\documentstyle[12pt]{article}
\newcommand{\no}{\nonumber\\}
\newcommand{\be}{\begin{equation}}
\newcommand{\ee}{\end{equation}}

\newcommand{\ba}{\begin{eqnarray}}
\newcommand{\ea}{\end{eqnarray}}
\newcommand{\ci}[1]{\cite{#1}}
\newcommand{\bi}[1]{\bibitem{#1}}
\newcommand{\la}[1]{\label{#1}}
\newcommand{\ce}[1]{\centerline{#1}}
\def\gl#1{(\ref{#1})}
\def\tr#1{\mbox{\rm tr}\left(#1\right)}

\begin{document}
%\begin{flushright}
%{\bf Preprint SPbU-IP-}\\
%\end{flushright}
\renewcommand{\thefootnote}{\fnsymbol{footnote}}
\vspace*{.5cm}

\begin{center}
{\Large\bf
The Extended Chiral Quark Model and QCD}
\end{center}

\vspace{1.0cm}

\ce{ A. A. Andrianov\footnotemark[1], D. Espriu\footnotemark[2] and
R.Tarrach\footnotemark[3]}
\medskip

\ce{\it Departament d'Estructura i Constituents de la Mat\`eria}
\ce{\it Universitat de Barcelona}
\ce{\it Diagonal, 647, 08028 Barcelona, Spain}

\vspace{2.0cm}
\ce{\sc ABSTRACT}
\medskip
\noindent We consider the low energy effective
action of QCD below the chiral symmetry breaking scale,
including, in Wilson's spirit,
all operators of dimensionality less or equal to 6 which can be built with
quark and chiral fields. The effect of the residual
gluon interactions
is contained in a number of coupling constants, whose running is studied.
The
resulting model is an extension of both the  chiral quark model
and the
Nambu-Jona-Lasinio one.
Constraints
on the coefficients of the effective lagrangian are
derived from the requirement of chiral symmetry restoration at
energies above the chiral symmetry breaking scale, from
matching to QCD at intermediate scales, and by fitting
some hadronic observables. In this model two types of
pseudoscalar states
(massless pions and massive $\Pi$-mesons), as well as one scalar
one arise as a consequence of dynamical chiral symmetry
breaking. Their masses and coupling constants
are studied. We also predict a number of low energy structural
constants. We find out that QCD favours a low-energy effective theory which
is largely dominated by the simplest chiral quark model, whereas
higher dimensional operators (such as those of the
Nambu-Jona-Lasinio type) can be assumed to be small. 

\vspace{1cm}
\noindent
PACS codes: \quad 12.38.-t,\quad 12.39.-x, \quad 14.40.-n\\

\noindent
Keywords:\quad Quantum Chromodynamics, Chiral Quark Models, Mesons\\
\vfill
\noindent
UB-ECM-PF 98/04\\
\noindent
February 1998\\
\noindent
hep-ph/9803232
\setcounter{footnote}{1}
\footnotetext{Permanent address:
Department of Theoretical
Physics, Sankt-Petersburg State University,
198904 Sankt-Petersburg,  Russia. E-mail: andrian@snoopy.phys.spbu.ru}

\setcounter{footnote}{2}
\footnotetext{
E-mail: espriu@ecm.ub.es}

\setcounter{footnote}{3}
\footnotetext{
E-mail: tarrach@ecm.ub.es}

\newpage
\setcounter{footnote}{0}
\renewcommand{\thefootnote}{\arabic{footnote}}
\noindent {\bf 1.  Introduction}

\medskip

Two QCD-inspired quark
models are commonly  used\footnote{A direct QCD bosonization approach
also exists \ci{AN}, which allows to construct light meson lagrangians
at low energies.} as  substitutes of
QCD in the hadronization regime[1-10].

The first one is the Chiral Quark Model (CQM) \ci{1}  with colored
quark fields $q_i^\alpha, \bar q_i^\alpha; i = 1,\ldots, N_c,  \alpha=1,2,..,N_F$
interacting with the colorless chiral field  $U(x)$,
by means of the Yukawa-type operator
\be
{\cal L}_{CQM} =
\bar q \left( i\!\not\!\partial
 -  M_0 (U P_L + U^+ P_R)\right) q
+ \frac14 f^2_0 \tr{\partial_{\mu}U\partial^{\mu}U^+},\la{CQM}
\ee
where   $ \not\!\partial \equiv \gamma^{\mu}
\partial_{\mu}$ and the projectors $P_{L(R)} = (1 \pm  \gamma_5)/2$ are
used.
$M_0$ is a chirally invariant constituent mass, which is assumed to acquire
a non-zero value once chiral symmetry is spontaneously broken.
$U(x) = \exp \left(i\pi(x)/ F_0\right)$  is a $SU(N_F)$ matrix
with generators $ \pi \equiv \pi^a T^a$, $a=1,..., N_F^2-1$.
This massless lagrangian \gl{CQM} is invariant under
$ SU_L(N_F) \bigotimes SU_R(N_F) $ symmetry  transformations of chiral
and quark fields
\be
 U \rightarrow \Omega_R U \Omega^+_L,\quad
q_L \rightarrow \Omega_L q_L,\quad
q_R \rightarrow   \Omega_R q_R,\la{trans}
\ee
and hence the fields $\pi(x)$ correspond to the massless Goldstone bosons.
For this paper we take $N_F=2$.

The constant $F_0$ is the pion  decay constant, whereas
$f_0$ represents a bare (as opposed to radiatively induced) pion decay
constant, which is supposed to contain
residual gluon contributions not accounted for by the constituent mass.
This bare $f_0$ is
considerably
enhanced by the radiative quark-loop  contribution to a final value
$F_0$. Setting $f_0=0$
is equivalent to assuming that the main
infrared effect of QCD ---the dynamical breaking of chiral symmetry--- is
associated solely to the appearance of the constituent mass $M_0$. On the
other hand, in a more precise treatment low-energy gluons
surely contribute in other ways and shift the
free-quark loop value of the pion constant. This effect is encoded in the
bare pion kinetic term\ci{2}.

The second approach is provided by  the Nambu-Jona-Lasinio model
(NJL)\ci{3}. It contains a
chirally invariant four-fermion interaction. For massless quarks it reads
\be
{\cal L}_{NJL} \, = \,i\bar q  \not\!\partial  q \,
+\,\frac{g}{4N_{c} \Lambda^2}\,
\left[(\bar q q )^{2} - (\bar q \gamma_5 \tau^a q )^{2}\right].
\la{NJL}
\ee
This lagrangian is also invariant under
$ SU_L(2) \bigotimes SU_R(2) $ symmetry  transformations
\gl{trans} of quark fields.
It is however well known
that for strong enough coupling  $g > g_c$
dynamical breaking of chiral symmetry occurs in this model. This
leads to the creation of a massless pion state, to the appearance
of a dynamical
quark mass $M_d$ (similar to $M_0$), and to the presence of a  massive
scalar meson state with the mass $m_{\sigma} \simeq 2M_d$, the sigma
particle.

Both models provide, after integrating out the fermions, basically
the same set of low energy structural constants describing interactions
between (pseudo) Goldstone bosons, in reasonable agreement
with data, specially for those operators which are quark mass independent.
However,
neither of these two models is really satisfactory at intermediate
energies $\sim 1 $ GeV. The CQM does not describe  scalar, pseudoscalar,
 etc. meson states with masses $\sim 1 $ GeV, which  are certainly
present in the
phenomenology\ci{10}. The NJL model (which can be extended to include
vector and axial-vector fields, in addition to pions and scalars\ci{6})
does not incorporate the heavy
pseudoscalar  $\Pi$ either\footnote{This meson is often identified with
the radial excitation of the pion and denoted as $\pi'$.}
with mass $m_{\Pi} \simeq 1.3$ GeV, thus treating this channel
asymmetrically.
Furthermore, its scalar sigma
meson is  too light to be identified with an observable
scalar quarkonium state. Both these models fail to correctly reproduce
the restoration of chiral symmetry at energies just above the chiral
breaking scale. This restoration is
 characteristic of QCD even at relatively low energies thanks
to the strong power suppression of the chirally non-invariant condensates.
For instance, the first QCD
contribution to the difference of scalar-pseudoscalar two-point
correlators is  of ${\cal O}(1/p^4)$ but even this one turns out 
to be negligible (see section 7).

Several ways have been  proposed \cite{2,6,7,7a} to modify the above
models in order
to bring them into
better agreement with phenomenology and to conform to the
predictions of QCD.
The quasilocal approach of \ci{7} (see also \ci{7a})
represents a generalization of the NJL model where four-fermion
operators with
derivatives are incorporated,  motivated by the soft momentum
expansion of the QCD effective action. For sufficiently strong couplings,
the new operators promote the formation of additional
scalar
and pseudoscalar states. These models allow an extension of the
linear $\sigma$ model provided by the NJL model,  with the pion being
a broken symmetry partner of the
lightest scalar meson just as before, and with excited pions and scalar
particles coming in pairs.

Clearly, to deal with the problem of which is the relevant
effective action of QCD at intermediate energies
we have to think in more general terms
and use only those constraints which can be derived
from symmetry and QCD principles and renormalization-group ideas.  We
propose here a wilsonian approach. We shall consider
the most general action describing
strong interactions once the spontaneous breakdown of
chiral symmetry has taken place at a scale $\Lambda$, without
committing ourselves to any specific mechanism or model.
This effective wilsonian action
should clearly contain the new light degrees of freedom
appearing below $\Lambda$, namely the (pseudo) Goldstone
bosons.
In this paper we thus adopt the point of view that
the pion
is a distinguished hadron state being the (pseudo) Goldstone
particle associated to the breaking of chiral symmetry. Their
couplings to fermions and to other mesons will therefore be the
ones indicated by the non-linear sigma model.
As, in the chiral limit, the pion is the only massless hadron state
one can presume that, after its explicit incorporation in the
low energy effective action, the remaining nonlocal operators
are expandable in powers of momenta
in all color-singlet (hadronic) channels.
We thus include all possible local operators up to dimension six (the
 choice of this upper dimension will be argued latter)
 which can be built with the help
of the chiral field $U$ and quarks  which are
gauge and chiral
invariant. Gluons themselves are not explicitly included. Rather the
effect of low energy
gluons is contained in a number of parameters, which will
depend on quantities like $\langle \alpha_s GG\rangle$ and similar
ones. These parameters cannot be determined at present and
they will be taken as given and used as boundary conditions
(at scale $\Lambda$) of our renormalization-group equations.
However, once properly normalized they
are expected to be of ${\cal O}(1)$ on naturalness grounds, unless
forbidden due to some QCD property (such as e.g. chiral symmetry).

Thus the effective lagrangian of strong interactions below the
breaking scale will be taken to be
\ba
{\cal L} &=&
\bar q \left( i\not\!\partial -
 M_0 (UP_L + U^+P_R)\right) q + \frac{1}{4} f_0^2
\tr{\partial_\mu U\partial^\mu U^\dagger}\no
&&+\Delta{\cal L}^{(4)}+
\Delta{\cal L}^{(5)}
+ \Delta{\cal L}^{(6)}. \la{hybr}
\ea
where we have made manifest the lowest dimensional operator
induced by the breaking. It is the one considered in the CQM
and we just remark here that is the one of lowest dimension
and hence expected to be the most relevant one. The remaining
operators have been classified according to their dimensionality
and lumped into
$\Delta{\cal L}^{(4)}$,
$\Delta{\cal L}^{(5)}$, etc. One of these operators is
the dimension 6 operator
\be
{g\over {4N_c\Lambda^2}}\left[
(\bar{q}q)^2 - (\bar{q}\gamma_5\vec{\tau} q)^2\right] ,
\la{opNJL}
\ee
but of course there are many more. Their role will be discussed
later.
The scale normalizing the 4-fermion coupling constant is supposed to be
related to the nonperturbative chiral symmetry breaking (CSB) scale
$\sim 1$ GeV.
Namely, we expect that attractive 4-quark forces are generated as a net
effect of high-momentum gluon exchange with the natural infrared cutoff
induced by the scale $\Lambda$. We have also
explicitly included the color factor $N_c$ to allow us to use the
large number of colors limit. In addition we work in the leading log
approximation. That is, we retain only those contributions which are
universal and regulator independent (sometimes they can be constant
when the logarithm is absent). Technically, this amounts to saying that
our results are exact in the limit where there is a large separation
between $M_0$ and
the chiral symmetry breaking scale $\Lambda$. While this is of course
not quite true in QCD, it is by no means a bad approximation and, in any
case, the error made is of the same order as the contributions
from operators which are beyond the
precision of our model anyway.

Let us now advance some of our results.
It will be proven that the mass spectrum of colorless states below $\Lambda$
(in the large-$N_c$ approach) includes two pseudoscalar states:
the pion,
massless in the chiral limit ($m_q = 0)$, and a heavy $\Pi$-meson,
and also one massive scalar state.
In our model  the $\Pi$-meson, rather than the pion,
is the natural partner of the scalar meson.
We have not done so explicitly, but the
model can be easily extended to include the lightest vector and axial-vector
QCD bound states. Keeping operators up to dimension six thus
provides a `one massive resonance plus continuum' model for the
spectral functions in the large $N_c$ limit. One could easily
include excited states by considering higher dimensional operators, but at
the price of pushing $\Lambda$ away from the chiral symmetry breaking
scale and making the meaning of the bare coefficients in the wilsonian
action less transparent. We thus treat excited states as part
of the continuum and integrate them out, as they are naturally heavier than
the chiral symmetry breaking scale. (Of course in large $N_c$
QCD we can adopt the opposite point of view, namely
to dispose entirely of fermions and gluons
and assume that the value of the constants of the effective
lagrangian at the chiral symmetry breaking scale $\Lambda$ is
provided by integrating the infinite tower of narrow resonances, or one
may take a position somewhere in between.)
Due to 4-quark forces the bare scalar
decay constant verifies $F_{\sigma} > F_{\pi}$
and the linear $\sigma$-model equality is no more valid for it.
However the residual chiral symmetry can be implemented in a way whose
restoration at high  energies is manifest. This residual invariance
provides generalized $\sigma$ model
relations between decay coupling constants for three kinds of mesons.
We will then see that phenomenology favors a weak coupling regime
for the dynamics associated to four-quark operators. QCD seems
to suggest that they play a marginal (albeit still phenomenologically
important) role in CSB.
We also derive the familiar chiral structural constants and
touch upon basically all relevant parameters in low energy
phenomenology.

The paper is structured as follows. In section 2 we formulate and motivate
the selection rules which are applied to build the most general
lagrangian. In section 3 we classify the operators
assembled in $\Delta{\cal L}^{(i)}$ according to their role in the
CSB, reduce their variety to a minimal set with
the help of the selection rules
and establish what we call
the Extended Chiral Quark Model (ECQM). In section 4 we
introduce in the ECQM the auxiliary scalar
and pseudoscalar fields in order to describe dynamical effects in the
large $N_c$ approach. The mass-gap equation
is derived and its solutions are
analyzed in the weak and strong coupling regimes.
In section 5 the mass-spectrum of scalar and
pseudoscalar fields is obtained and the pion decay constant is determined
in terms of the bare parameters of the model. In section 6  external vector,
axial-vector, scalar and
pseudoscalar sources are attached to the model. By these
means the  decay coupling constant of the pion is calculated and
the current quark mass $m_q$ is incorporated. The
decay constant of the heavy  $\Pi$ is also estimated
to first order in  $m_q$.
In section 7 a large $N_c$  analysis of two-point correlators in QCD is
performed and chiral symmetry restoration is analyzed
on the basis of the operator product expansion (OPE)
for high Euclidean momenta.
The condition of chiral symmetry restoration
leads to two sum rules \`a la Weinberg, but in the scalar channel.
In section 8 the fitting of parameters of the ECQM is performed with
the help of a number of observables and theoretical inputs.
Finally section 9  contains our conclusions. Some technical details
as well as some considerations which are not along the main line
of the paper
have been relegated to the appendices.

\bigskip

\noindent{\bf 2. Selection rules}

\medskip

In this section we shall
present the list of the rules allowing us to select
the effective lagrangian operators
and parameters
of the Extended Chiral Quark Model. They are the following
\begin{enumerate}
\item Locality.
\item $P$  and $C$ invariance of the Lagrangian.
\item Chiral and gauge invariance.
\item Chiral Perturbation Theory  (ChPT) counting rules for chiral
fields at low energies, and the
canonical dimension counting rules for quark fields valid
for intermediate energies.
\item The incorporation of all operators
of dimensionality smaller or equal to
six, in order to take into consideration
 the qualitatively new
dynamics originating from the local four-quark interactions.
\item The large $N_c$ limit.
\item The assumption that residual gluon interactions
below the chiral symmetry breaking (interactions whose strength must,
logically, be much diminished, as the pion
carries a substantial part of the strong force) can be included as bare
coefficients of the effective lagrangian at scale $\Lambda$. These
contributions will be deemed small at this scale
and hence they will
only be considered for the leading operators in the chiral lagrangian.
\item The subsequent neglect of bare operators of
next-to-leading order in ChPT which
contain four or more derivatives acting on mesons and no quark fields.
We shall assume that the
bulk of the contribution to these operators
will be the one induced radiatively after
integration of the fermions. (For explicit expressions of the bare
coefficients of the Gasser-Leutwyler lagrangian
in terms of the gluon condensate, both for the leading
and next-to-leading terms, see \ci{2}.) In any case
the eventual inclusion of non-zero  bare values for the
coefficients for the
${\cal O}(p^4)$ and ${\cal O}(p^6)$ chiral operators
adds only to the value of the Gasser-Leutwyler coefficients
and is of little consequence for the rest of the analysis.
\item The large cutoff approximation
in the radiatively induced
bosonic operators with the cutoff $\Lambda$ being the
scale of the
chiral symmetry breaking. It will be applied in a strong
sense
as a large log approximation in the cutoff dependence which consistently
supports the neglect of operators of dimension higher than 6 (see \ci{7}).
\item The neglect
of color singlet higher spin current-current
interactions whose description requires
explicit vector, axial-vector and
higher-spin bosons. Likewise, we restrict ourselves to the study of
isosinglet
scalar and isotriplet pseudoscalar meson physics only. The
extension to other fields is straightforward. This restriction
does not affect the accuracy of the model other than by
shifting the bare value of $g_A$.
\item The use of the equations of motion (EOM)
of the CQM (with minimal number of
derivatives) in order to replace
derivatives in operators with the mass-like parameters and  with
higher-order multi-quark and multi-pion interactions.
\item The restoration of chiral symmetry
at energies higher than $\Lambda$,
that is,  the fact that scalar current correlators should approach
the pseudoscalar ones with relative asymptotics not worse than
in QCD.
\item The normalization of the resulting pion decay constant $F_{\pi}$
to the phenomenologically observed value and of the quark condensate to the
typical theoretical input
 $\langle \bar q q\rangle = - (200\div 250 {\rm MeV})^3$.
\item
The chiral dynamics relation
\be
\Pi_S (0) - \bar\Pi_P^{aa}(0) = 64 L_{8}\frac{\langle\bar q
q\rangle^2}{F^4_{\pi}} \, , \la{L8}
\ee
where $\bar\Pi_P^{aa}(0)$ stands for the zero-momentum value of
the pseudoscalar two-point correlator
with subtracted
pion pole term projected onto the isospin singlet channel, and $L_8$ is a
chiral constant  of the Gasser-Leutwyler Lagrangian \ci{13}.
\end{enumerate}

\bigskip

\noindent{\bf 3. Complete set of operators of $d\leq 6$ in the ECQM}

\medskip

The complete set of chirally symmetric operators of dimension less or
equal to six
which in principle should be included in the effective Lagrangian
can be divided into the following subsets: (a) purely chiral operators;
(b),(c),(d),(e) operators bilinear in the fermion fields of
$d=3,4,5$ and 6, respectively;
(f) four-fermion interactions of $d=6$.
For the description of the above operators it is convenient to
rewrite the derivative of the chiral field in terms  of
chiral currents
\be
l_\mu \equiv  U^\dagger \partial_\mu U,\qquad
r_\mu \equiv - (\partial_\mu U)  U^\dagger\, .\la{chcur}
\ee

\medskip
\noindent (a) The purely chiral operators
relevant for the description of low-energy
pion dynamics include some with two chiral currents
\ba
&&\tr{ l_\mu  l^\mu} =
- \tr{ \partial_\mu U \partial^\mu U^\dagger},
 \la{ch22} \\
&&\tr{ \partial_\mu l^\mu \partial_\nu l^\nu}, \la{ch24} \\
&&\tr{ \partial_\mu\left(\partial_\nu l^\nu\right) \partial^\mu
\left(\partial_\sigma l^\sigma\right)}. \la{ch26}
\ea
Similar operators with $l_\mu \to r_\mu$ are in fact not independent.
In addition we have operators with four or six chiral
currents.
(See
\ci{13} and  \ci{14} for a complete list.)
They give next-to-leading
contributions to strong interaction processes in the derivative
expansion, namely their contribution
contains at least four powers of external momenta.
As discussed in section 2, we shall neglect these operators
assuming that at the scale of chiral symmetry breaking $\Lambda$
the corresponding bare coefficients are small.
The operators
are written in a form which is specially convenient in order
to use
the EOM for the chiral field. These we derive, as usual, from the
lowest dimensional effective lagrangian, namely the CQM
in (\ref{CQM}), and they read
\ba
(\partial_\mu l^\mu)_{ij} &=& \frac{2 M_0}{f^2_0}\left[
(\bar q_{R} U)_i q_{Lj} - \bar q_{Lj} (U^\dagger q_{R})_i
- \frac12 \delta_{ij} (\bar q_R U q_L -
\bar q_L U^\dagger q_R)\right]\no
&=& \frac{M_0}{f^2_0} (\vec\tau)_{ij}
\left[\bar q_R U\vec\tau q_L -
\bar q_L\vec\tau U^\dagger q_R \right].
\la{em1}
\ea
A similar equation\footnote{The last term in the first line of
\gl{em1} appears for   $SU(2)$
chiral fields and is
absent for $U(2)$ fields.}
can obtained for the
right current $r_\mu$ with the replacement $U  \rightarrow  U^\dagger;\quad
l_\mu \rightarrow  r_\mu;\quad L \leftrightarrow R$.
Eqs. \gl{em1} allow us to replace operators \gl{ch24}, \gl{ch26}
by four-fermion ones.

\medskip
\noindent (b) This class is particularly simple. There is only one
operator of $d = 3$ and it is a chirality flipping one
\be
\bar{q}_R U q_L + \mbox{\rm h.c.}\, . \la{d3}
\ee

\medskip
\noindent (c) The operators of $d = 4$ are chirality preserving
\ba
&&\bar{q}_L\not\!\partial\, q_L +
\bar{q}_R\not\!\partial \,q_R , \la{d41}\\
&&\bar{q}_L \not l \, q_L +
\bar{q}_R \not r \, q_R . \la{d42}
\ea
The latter operator contributes to the $g_A$ constant of
the (constituent) quark \ci{1}.

\medskip
\noindent (d) The complete set of $d=5$ operators, which are
chirality flipping, can be built from
operators with a `normally ordered' differential structure
(i.e. all the derivatives either act on chiral fields or to the right,
 on quark fields $q_L, q_R$). In particular,
the full list of these reads\footnote{In h.c.
terms $U^\dagger$ and right chiral currents $r_\mu$
appear in a mirror reflected way. Therefore
all the analysis done with left currents is valid also for the
h.c. terms.}
\ba
&&\bar{q}_R U \partial^2 q_L +\mbox{\rm h.c.},
 \la{d51}\\
&&\bar{q}_R U (\partial_\mu l^\mu) q_L +\mbox{\rm h.c.},
 \la{d52}\\
&&\bar{q}_R U  l^\mu  \partial_\mu q_L +
\mbox{\rm h.c.}, \la{d53}\\
&&\bar{q}_R U l_\mu  l^\mu q_L +
\mbox{\rm h.c.}, \la{d54}\\
&&\bar{q}_R U \sigma^{\mu\nu} l_\mu \partial_\nu q_L +
\mbox{\rm h.c.}, \la{d55}\\
&&\bar{q}_R U \sigma^{\mu\nu} [l_\mu,  l_\nu] q_L +
\mbox{\rm h.c.}, \la{d56}\\
&&\bar{q}_R U \sigma^{\mu\nu} (\partial_\mu l_\nu) q_L +
\mbox{\rm h.c.}, \la{d57}
\ea
where $\sigma^{\mu\nu} \equiv [\gamma^\mu, \gamma^\nu]$.
These seven operators are not independent however. Namely, the identity
\be
\partial_\mu l_\nu - \partial_\nu l_\mu +  [l_\mu,  l_\nu] = 0 \la{id1}
\ee
allows to replace (\ref{d57}) in favour of (\ref{d56})
and one is left with 6 independent operators.
A further reduction is provided by applying the  EOM, which we
again borrow
from the lowest dimensional CQM (\ref{CQM})
\be
i\not\!\partial q_L =  M_0 U^+ q_R,\qquad
i\not\!\partial q_R =  M_0 U q_L .\la{em2}
\ee
See the appendix A for a justification of our usage
of the EOM.

In order to use the simplification brought about by the EOM
most effectively it is convenient  to rewrite all the
operators in such a way that the Dirac derivative acts either on
the left or on the
right side. Two independent differential
identities are useful for this purpose
\ba
&&U \not{l} \,\not\!\partial\, =\,  U  l^\mu  ( \partial_\mu +
\frac12  U \sigma^{\mu\nu} l_\mu \partial_\nu), \la{id2}\\
&&\not\!\partial \, U \!\not{l} \,\,+\, U \not{l} \,\not\!\partial\, =\,
2  U  l^\mu  \partial_\mu +  U (\partial_\mu l^\mu) +
 U l_\mu  \l^\mu\, . \la{id3}
\ea
Then any two of the operators (\ref{d53}),
(\ref{d54}), (\ref{d55}) can be replaced with a combination of the third one
and the operators
\ba
&&\bar{q}_R U \!\not{l} \,(\not\!\partial   q_L) +
\mbox{\rm h.c.},\la{d58}\\
&&\bar{q}_R \not\!\partial ( U \!\!\not{l} \, q_L ) +
\mbox{\rm h.c.}, \la{d59}
\ea
which in turn can be related to already considered lower dimensional
operators  after application
of the EOM. Similarly, the first two operators (\ref{d51}), (\ref{d52})
can be subject to EOM reduction
with the help of (\ref{em1}),(\ref{em2}).
Finally we end up with only two relevant operators of $d=5$.
Let us take the operators (\ref{d54}) and  (\ref{d56}) as the basic ones.

Finally, we analyze the structure of (\ref{d56}) and discover that
at low energies, when $l_\mu \simeq i \partial_\mu \pi^a \tau^a$,
this operator becomes
\be
\bar{q}_R U \sigma^{\mu\nu} [l_\mu,  l_\nu] q_L +
\mbox{\rm h.c.} \simeq - 2i \epsilon^{abc}
\partial_\mu \pi^a \partial_\nu \pi^b
\bar{q}_R \sigma^{\mu\nu}\tau^c  q_L +
\mbox{\rm h.c.}.  \la{le1}
\ee
The quark bilinear in the above equation, being proportional to the
antisymmetric
tensor current, will be interpolated after bosonization by derivatives
acting on bosonic fields:
$\epsilon^{abc}
\partial_\mu \Phi^a \partial_\nu \Phi^b $. It will involve
either two pseudoscalar fields
or two scalar ones as we do not consider any vector or tensor
fields in this paper. Therefore
after integration of the fermion fields, only
terms with four or more derivatives acting on scalar fields will be
produced by this operator.
Thus (\ref{d56}) is on the same footing as the next-to-leading set of
operators containing at least
four derivatives of bosonic fields, whose bare coefficients
we agreed not to consider.
In fact, on dimensional grounds based on our selection rules, the
present contribution will be suppressed
with respect to the purely
bare value of the ${\cal O}(p^4)$ operators by powers of $M_0/\Lambda$
and therefore is subleading even with respect to the bare value
of the ${\cal O}(p^4)$ coefficients, which we
anyway neglect.

With these arguments only one operator, say  (\ref{d54}), should be
taken into account in the ECQM.

\medskip
\noindent (e) We move next to the $d=6$ operators which can be built out
of a quark bilinear by adding derivatives and chiral currents.
This category consists
of 25 (chirality preserving)
operators with three derivatives,  of course not all of them independent
\be
\bar{q}_L \hat{\cal O}^L_{3} q_L +
\bar{q}_R \hat{\cal O}^R_{3} q_R ,\la{v6}
\ee
where the `normal ordered' differential operator $\hat{\cal O}^L_{3}$
can be any from the following list \footnote{The operators
$\hat{\cal O}^R_{3}$ are obtained from
$\hat{\cal O}^L_{3}$ by replacement $l_\mu \longrightarrow r_\mu$.}
\ba
&&\not\!\partial \,\partial^2,\\
&& \not{l} \, \partial^2,\qquad\partial^2(\!\!\not{l}),
\qquad(\partial_\mu l^\mu)\not\!\partial,\qquad
 l^\mu  \partial_\mu \not\!\partial,\\
&&\not\!\partial(l_\mu)\partial^\mu,\qquad
(\partial_\mu \!\!\not{l})\partial^\mu,\qquad
 \not\!\partial(\partial_\mu l^\mu) \qquad
\epsilon_{\mu\nu\rho\lambda}\gamma_5 \gamma^\mu  (\partial^\nu  l^\rho)
\partial^\lambda,\\
&&l_\mu l^\mu  \not\!\partial,\qquad
\not\!\partial(l_\mu) l^\mu,\qquad
l^\mu \not\!\partial(l_\mu),\qquad l^\mu (\partial_\mu \!\!\not{l}),
\qquad (\partial_\mu \!\!\not{l})l^\mu,\\
&&\not l (\partial_\mu l^\mu),\qquad (\partial_\mu l^\mu) \!\!\not l,
\qquad l_\mu \not{l} \partial^\mu,\qquad \not{l} l_\mu \partial^\mu,
\qquad \epsilon_{\mu\nu\rho\lambda}\gamma_5 \gamma^\mu  l^\nu  l^\rho
\partial^\lambda,\\
&&\epsilon_{\mu\nu\rho\lambda}\gamma_5 \gamma^\mu
 (\partial^\nu  l^\rho)  l^\lambda, \qquad
\epsilon_{\mu\nu\rho\lambda}\gamma_5 \gamma^\mu
 l^\nu  (\partial^\rho  l^\lambda), \\
&&\epsilon_{\mu\nu\rho\lambda}\gamma_5 \gamma^\mu
 l^\nu  l^\rho  l^\lambda, \qquad
 \not l \,l_\mu  l^\mu,\qquad  l_\mu\!\!\not l \, l^\mu,\qquad
 l_\mu  l^\mu \!\!\not l . \la{full}
\ea
We use the conventions $\gamma_5 = i\gamma_0\gamma_1\gamma_2 \gamma_3$
and $\epsilon _{0123} = 1$.
Not all of them are independent, of course. We have the following
seven identities
\ba
&&\partial^2(\!\!\not{l}) = \not\!\partial(\partial_\mu l^\mu)
+ [\not l, (\partial_\mu l^\mu)] - [ l^\mu,  (\partial_\mu \!\!\not{l})],
\la{d64}\\
&&(\partial_\mu \!\!\not{l})\partial^\mu =
[\not\!\partial, l_\mu  \partial^\mu]
-  [l_\mu,\, \not{l}]\, \partial^\mu, \la{d68}\\
&&\{l^\mu, (\partial_\mu \!\!\not{l})\} = [\not\!\partial, l_\mu l^\mu]
+ \{l^\mu, [\!\!\not{l},l_\mu] \},\la{d65}\\
&&[l^\mu, \not\!\partial(l_\mu)] = [l^\mu, (\partial_\mu \!\!\not{l})]
- [l^\mu,[ l_\mu, \not l]] ,\la{d66}\\
&&\epsilon_{\mu\nu\rho\lambda}\gamma_5 \gamma^\mu  (\partial^\nu  l^\rho)
\partial^\lambda =
- \epsilon_{\mu\nu\rho\lambda}\gamma_5 \gamma^\mu  l^\nu  l^\rho
\partial^\lambda,\\
&&\epsilon_{\mu\nu\rho\lambda}\gamma_5 \gamma^\mu
 (\partial^\nu  l^\rho)  l^\lambda =
- \epsilon_{\mu\nu\rho\lambda}\gamma_5 \gamma^\mu
 l^\nu  l^\rho  l^\lambda
 = \epsilon_{\mu\nu\rho\lambda}\gamma_5 \gamma^\mu
 l^\nu  (\partial^\rho  l^\lambda)
\ea
and one is left with 18 independent operators in
\gl{v6}.

Of these, ten operators are subject to obvious reduction
with the help of the EOM. They are the ones containing
\ba
&&\not\!\partial \,\partial^2,\qquad \not{l} \, \partial^2,
\qquad(\partial_\mu l^\mu)\not\!\partial,\qquad
 l^\mu  \partial_\mu \not\!\partial,\qquad
l_\mu l^\mu  \not\!\partial,\la{d61}\\
&& [\not\!\partial, l_\mu \partial^\mu] =
\not\!\partial(l_\mu)\partial^\mu,\qquad
[\not\!\partial, l_\mu l^\mu] = \{\not\!\partial(l_\mu), l^\mu\} ,\la{d62}\\
&& [\not\!\partial , (\partial_\mu l^\mu)],
\qquad \not{l}\, (\partial_\mu l^\mu),
\qquad (\partial_\mu l^\mu) \!\!\not l.  \la{d63}
\ea
Evidently the related 10 operators in \gl{v6}
are eliminated straightforwardly. Of the remaining
ones, three operators, namely
\ba
&&[l^\mu, (\partial_\mu \!\!\not{l})],\qquad
[l_\mu,\, \not{l}]\, \partial^\mu,\la{d69}\\
&&\epsilon_{\mu\nu\rho\lambda}\gamma_5 \gamma^\mu  l^\nu  l^\rho
\partial^\lambda, \la{d611}
\ea
are involved in the following two identities
\ba
&&2 [l_\mu,\, \not{l}] \partial^\mu + [l^\mu, (\partial_\mu \!\!\not{l})]
  = [\not{l}, \partial^2]  + [\not\!\partial, (\partial_\mu l^\mu)]
 + 2 [\not\!\partial, l_\mu \partial^\mu] +
 [\!\!\not l , (\partial_\mu l^\mu)],
\la{id4}\\
&&i \epsilon_{\mu\nu\rho\lambda}\gamma_5 \gamma^\mu  l^\nu  l^\rho
\partial^\lambda
- [l_\mu,\, \not{l}]\, \partial^\mu = \frac12 \sigma_{\mu\nu} l^\mu l^\nu
\not\!\partial,
\la{id5}
\ea
and, therefore, by using the EOM one
can discard the operators \gl{d69} in favour
of \gl{d611}.

Finally, one is left with the following 6 operators:
\be\{\not{l},\, l_\mu\} \partial^\mu
  - \frac12 l_\mu l^\mu \not\!\partial ,
\la{d610}
\ee
$$
\epsilon_{\mu\nu\rho\lambda}\gamma_5 \gamma^\mu  l^\nu  l^\rho
\partial^\lambda, \la{d6110}$$
\be\epsilon_{\mu\nu\rho\lambda}\gamma_5 \gamma^\mu
 l^\nu  l^\rho  l^\lambda, \qquad
 \not l \,l_\mu  l^\mu,\qquad  l_\mu\!\!\not l \, l^\mu,\qquad
 l_\mu  l^\mu \!\!\not l . \la{d612}
\ee
Notice that one of the operators contains a piece which is
manifestly reducible by the use of the EOM. We have kept it
to make the associated quark bilinear traceless. Note also that
the analysis of the structure of other possible EOM-reducible
operators with a
different position of the Dirac derivative
$\not\!\partial$ does not yield any new constraints.

We notice that the above structures contain at least
two chiral currents. Let us consider, for instance,
the quark bilinear
containing an insertion of the first of the above structures.
Since
$l_\mu \simeq i \partial_\mu \pi^a \tau^a$,
\be
\bar q_L\{\not{l},\, l_\mu\} \partial^\mu
 q_L \simeq
\left( - 2 \partial_\mu \pi^a \partial_\nu \pi^a
+ \frac 12 g_{\mu\nu} \partial_\rho \pi^a \partial^\rho \pi^a\right)
\bar q_L \gamma^\mu \partial^\nu q_L. \la{le2}
\ee
Since we neglect tensor fields, at low energies
the fermion bilinear
$\bar q_L \gamma^\mu \partial^\nu q_L$ must be bosonized
as
\ba
&& A \left(\partial^\mu \partial^\nu -
\frac14 g^{\mu\nu} \partial^2\right )
S
+ B \left( \partial^\mu P \partial^\nu P
- \frac 14 g^{\mu\nu} \partial_\rho P \partial^\rho P\right)\no
&&+ C \left( \partial^\mu S \partial^\nu S
- \frac 14 g^{\mu\nu} \partial_\rho S \partial^\rho S \right)
\la{le3}
\ea
where $S, P$ stand here for generic scalar and
pseudoscalar fields
and $A,B,C$ are bosonization constants.
As a result all fermion bilinear $d=6$ operators lead, after
integration of the fermion fields, to ${\cal O}(p^4)$
terms in the low energy effective chiral lagrangian, which,
just as before, we need not consider as they are suppressed
by a factor of ${\cal O}(M_0/\Lambda)$ with respect to the bare
${\cal O}(p^4)$ terms which we have already neglected.
Therefore there are no operators
of type (e)
which are eligible to be included in the leading wilsonian
action.

Insofar the effective operators considered are leading in the
large $N_c$ expansion.
To complete the list of chirally invariant operators bilinear in quark
fields one should mention also those ones built of two disconnected
traces in isospin indices and according to the QCD-motivated
combinatorial estimations \ci{15} they may arise in the next-to-leading
 large-$N_c$ order as a result of chiral bosonization involving
two quark loops. For $SU(N)$ chiral fields\footnote{If the $U_A(1)$ extension
of the ECQM were considered one would have to add 32 more independent
operators. However in this paper we adopt the viewpoint that the
$U(1)$ pseudoscalar boson is not chiral as it remains massive in the chiral
limit. Therefore it should arise from an effective quark selfinteraction.}
these operators are
\ba
&&\bar{q}_R U q_L \tr{l_\mu  l^\mu}  +
\mbox{\rm h.c.}, \la{n1}\\
&&\bar q_L\gamma_\mu \partial_\nu q_L  \tr{l^\mu  l^\nu}
 - \frac14 \bar q_L \not\!\partial q_L  \tr{l_\mu l^\mu} +
(L, l_\mu \to R, r_\mu), \la{n2}\\
&&\bar q_L\gamma_\mu l^\mu q_L  \tr{l_\nu  l^\nu}
- (L, l_\mu \to R, r_\mu), \la{n3}\\
&&\bar q_L\gamma_\mu l_\nu q_L  \tr{l^\mu  l^\nu}
- (L, l_\mu \to R, r_\mu), \la{n4}\\
&&\bar q_L\gamma_\mu  q_L \tr{l^\mu l_\nu l^\nu}
- (L, l_\mu \to R, r_\mu), \la{n5}\\
&&\epsilon_{\mu\nu\rho\lambda} \bar q_L \gamma^\mu  q_L
\tr{ l^\nu  l^\rho  l^\lambda} - (L, l_\mu \to R, r_\mu) \la{n6}\\
&&= - \epsilon_{\mu\nu\rho\lambda} \bar q_L \gamma^\mu  q_L
\tr{ (\partial^\nu  l^\rho) l^\lambda} - (L, l_\mu \to R, r_\mu), \la{n7}\\
&&\bar q_L \not\!\partial q_L  \tr{l_\mu l^\mu}
+ (L, l_\mu \to R, r_\mu),\la{n8}\\
&& \bar q_L \gamma^\mu  q_L  \tr{ l_\mu  (\partial_\nu  l^\nu)}
+ (L, l_\mu \to R, r_\mu), \la{n9}\\
&& \bar q_L \gamma^\mu  q_L  \tr{ l^\nu  (\partial_\nu  l_\mu)}
+ (L, l_\mu \to R, r_\mu) \la{n10a}\\
&& = \bar q_L \gamma^\mu  q_L  \tr{ l_\nu  (\partial_\mu  l^\nu)}
+ (L, l_\mu \to R, r_\mu)\no
&& = - \frac12
\partial_\mu(\bar q_L \gamma^\mu  q_L)  \tr{ l_\nu l^\nu} +
(L, l_\mu \to R, r_\mu) . \la{n10}
\ea
The operator \gl{n7} is reducible due to \gl{id1}.
Three operators \gl{n8}, \gl{n9}, \gl{n10a}, \gl{n10} can be
eliminated by means of the EOM.
The operator \gl{n5} vanishes in the $SU(2)$ case but may contribute for
any chiral symmetry group with nontrivial symmetric structure
constants, in particular, for $U(2)$ and $SU(3)$ chiral groups.
An analysis similar to the one already performed
leads to the conclusion that all the operators \gl{n2} -- \gl{n6}
contain at least four powers of momenta (four derivatives) and should,
to be consistent with our selection rules, be neglected.
Thus only one operator, \gl{n1}, should be included in the leading-order
lagrangian. But for the $SU(2)$ case
it coincides with \gl{d54} and does not
introduce any new phenomenological constant.

\medskip
\noindent (f) We finally turn to $d=6$ four-fermion operators.
Amongst them, according to our selection rules, we retain only those built
from color singlet scalar and pseudoscalar quark densities
supplemented with chiral fields.
We shall choose a basis for these operators consisting in
bilinears with
a definite isospin in the color-singlet channel
\ba
&&\Phi_1 = \bar{q}_L U^\dagger q_R \bar{q}_R  U q_L,  \la{4f1}\\
&&\Phi_2 = \bar{q}_L \tau^a U^\dagger q_R \bar{q}_R  U \tau^a q_L,  \la{4f2}\\
&&\Phi_3 = \bar{q}_L U^\dagger q_R \bar{q}_L U^\dagger q_R +
\bar{q}_R U q_L \bar{q}_R U q_L,
\la{4f3} \\
&&\Phi_4 = \bar{q}_L \tau^a U^\dagger q_R \bar{q}_L \tau^a U^\dagger q_R +
\bar{q}_R U \tau^a q_L \bar{q}_R U \tau^a  q_L. \la{4f4}
\ea
where color indices are contracted within each quark bilinear.

Depending on the choice of the chiral group, there are a few invariant
combinations which do not depend manifestly on chiral fields.
In the particular case of
$SU(2)$ there are two invariants
\ba
&&(\bar q q)^2 - (\bar q \gamma_5 \vec\tau q)^2 = 2\Phi_1 +  2\Phi_2
+ \Phi_3 - \Phi_4; \la{chi1}\\
&&(\bar q \gamma_5 q)^2 - (\bar q \vec\tau q)^2 = - 2\Phi_1 -  2\Phi_2
+ \Phi_3 - \Phi_4, \la{chi2}
\ea
while in general only their difference is invariant and $U$-independent
\be
2 \bar{q}_L^i q_R^j \bar{q}_R^j q_L^i =  \Phi_1 +  \Phi_2 . \la{chi3}
\ee

As discussed in section 2, for simplicity we shall only
consider the isosinglet scalar and isovector pseudoscalar
channels. (The generalization to other channels is straightforward
and it does not modify in any essential way the results presented
in the coming sections.) Then, of all the four fermion
operators in
\gl{4f1} - \gl{4f4}, we need to take into consideration the following
combinations
\ba
&&\Phi_{S0} =  (\bar{q}_L U^\dagger q_R  +
\bar{q}_R  U q_L)^2 =  2\Phi_1 +  \Phi_3,
\la{min1}\\
&&\Phi_{P1} = ( -  \bar{q}_L \vec\tau U^\dagger q_R
+  \bar{q}_R  U \vec\tau q_L)^2 =
 \Phi_4 -  2\Phi_2. \la{min2}
\ea
There are theoretical
arguments (see appendix B) suggesting that the coefficients
of these operators,
$g_{S0}$
and  $g_{P1}$,
 must be identical at the chiral symmetry
breaking scale $\Lambda$. However
these operators
may acquire different coupling constants below
the chiral symmetry breaking scale due to
the use of the EOM on other operators of $d\le 6$.
Indeed, the reduction of the dim-6 operator \gl{ch24} and
of the dim-5 operator \gl{d52} with the help of the EOM \gl{em1}
yields a contribution into the operator \gl{min2} changing
the value of  $g_{P1}$. Thus, even if the constants   $g_{S0}$
and  $g_{P1}$ are initially saturated by equal high energy gluon
exchange (as chiral symmetry holds in QCD perturbation theory)
they are split after the EOM are applied.
We shall keep them as independent coupling constants.

\bigskip

\noindent{\bf 4. Minimal ECQM and mass-gap equation }

\medskip

We shall from now on work in  euclidean space-time, making the
appropriate Wick rotation back to Minkowski space when required,
and regularize the
effective action obtained after integration of the fermions
by using a manifestly chiral invariant
cutoff. Let us now put together all the operators which
have survived the previous analysis.
In the chiral limit we have
\ba
{\cal L}_{ECQM} &=&
i\bar q \left( \not\!\! D_A
 +   M_0 (UP_L + U^+P_R)\right) q - \frac{f_0^2}{4}\tr{ l_\mu  l_\mu}\no
&&+\,\frac{g_{S0}}{4N_{c} \Lambda^2}\,
(\bar{q}_L U^\dagger q_R  +
\bar{q}_R  U q_L)^2 - \frac{g_{P1}}{4N_{c} \Lambda^2}\,
( -  \bar{q}_L \vec\tau U^\dagger q_R
+  \bar{q}_R  U \vec\tau q_L)^2 \no
&& + i \frac{\delta f_0}{\Lambda^2} \left[\bar{q}_R U l_\mu  l_\mu q_L +
\bar{q}_L U^\dagger r_\mu  r_\mu q_R\right],
\la{ECQM}
\ea
where
\be
\not\!\! D_A \equiv \not\!\partial  +  \frac12 \delta g_A (\!\not l P_L +
\!\not r P_R) .
\ee
As just discussed, $g_{S0}$ and $g_{P1}$ need not be equal.
The reason to call the coupling constant in the
last operator $\delta f_0$ is that it leads, after integration of the fermions,
to a redefinition of $f_0$.
This operator is thus inessential in the
analysis of two point correlators in the chiral limit
and becomes significant only by providing
a contribution of ${\cal O}(p^4)$ and beyond. This contribution is
suppressed by a power of $M_0^2/\Lambda^2$ with respect to the bare value,
which we consistently neglect anyway. Away from the chiral
limit $\delta f_0$ should be kept, however,
as we will discuss in section 6.
At this point we notice something remarkable. If we ignore all bare
${\cal O}(p^4)$ contributions to the low energy effective action,
the most general low energy action looks like a hybrid between the
Chiral Quark and the Nambu-Jona-Lasinio models. However, as soon
as we are interested in contributions of higher order
in the momentum expansion, a number (not too large, fortunately)
of bare couplings do contribute, modifying the
dominant contribution from the fermion loop.

In order to simplify further the calculations
it is helpful to introduce the fields
\be
Q_L \equiv \xi q_L,\qquad Q_R \equiv \xi^\dagger q_R,\qquad  \xi^2\equiv U,
\la{chv}
\ee
Under chiral transformations
\be
\xi \longrightarrow h_{\xi} \xi \Omega^+_{L} = \Omega_R \xi h^+_{\xi}.
\la{nlr}
\ee
In these variables the ECQM action reads
\ba
{\cal L}_{ECQM} &=&
i\bar Q \left( \not\!\! D
 +   M_0 \right) Q - \frac{f_0^2}{4}\tr{ a_\mu  a_\mu}\no
&&+\,\frac{g_{S0}}{4N_{c} \Lambda^2}\,
(\bar{Q}_L Q_R  +
\bar{Q}_R Q_L)^2 - \frac{g_{P1}}{4N_{c} \Lambda^2}\,
( -  \bar{Q}_L \vec\tau Q_R
+  \bar{Q}_R  \vec\tau Q_L)^2 \no
&& + i \frac{\delta f_0}{\Lambda^2} \bar{Q} a_\mu  a_\mu Q ,
\la{ECQM1}
\ea
where
\ba
&& Q \equiv Q_L + Q_R,\no
&&\not\!\! D \equiv \not\!\partial  + \frac12 \not\! v +
\frac{1}{2} \gamma_5 g_A \not\! a , \qquad g_A \equiv 1 - \delta g_A,
\no
&& v_\mu \equiv \xi^\dagger(\partial_\mu\xi) -
(\partial_\mu\xi) \xi^\dagger,
\qquad
a_\mu \equiv \xi^\dagger(\partial_\mu\xi) + (\partial_\mu\xi) \xi^\dagger.
\la{va}
\ea

Let us now introduce
auxiliary
scalar, $\widetilde\Sigma$, and pseudoscalar, $\widetilde\Pi =
 \widetilde\Pi^a \tau^a $, fields in order to
make the lagrangian
(\ref{ECQM}) bilinear in fermion variables. We can then
replace
\be
\frac{g_{S0}}{4N_{c} \Lambda^2}\,
(\bar{Q}_L  Q_R  +
\bar{Q}_R Q_L)^2 - \frac{g_{P1}}{4N_{c}\Lambda^2}\,
( -  \bar{Q}_L \vec\tau Q_R
+  \bar{Q}_R \vec\tau Q_L)^2
\ee
by
\be
i \bar{Q}_R ( \widetilde\Sigma (x)
+ i\widetilde\Pi(x)) Q_L +
i \bar{Q}_L ( \widetilde\Sigma (x)
- i\widetilde\Pi(x)) Q_R
+ N_{c} \Lambda^2\left(\frac{\widetilde\Sigma (x)^2}{g_{S0}} +
\frac{(\widetilde\Pi^a(x))^2}{g_{P1}}\right) \la{HS}
\ee
and an integration over the auxiliary variables.
The tildes on the fields $\widetilde\Sigma$ and
$\widetilde\Pi$ remind
us that they are for the time being just auxiliary variables.
The shift
\be
\widetilde\Sigma + i \widetilde\Pi
=  H(x) - M_0  -
\frac{\delta f_0}{\Lambda^2} a_\mu  a_\mu, \quad
 H(x) \equiv \Sigma + i \widetilde\Pi, \la{shi}
\ee
further simplifies the dependence
on chiral fields
and eventually one comes to the following lagrangian
\ba
{\cal L}_{ECQM} &=&
i\bar Q \left( \not\! D
 +  H(x)P_L + H^+(x) P_R\right) Q +
{\cal L}_{bos}, \no
{\cal L}_{bos} &\equiv&
\frac{N_c \Lambda^2}{8}\,
\tr{\widehat H(x)^T \widehat g^{-1} \widehat H(x) }
- \frac{ f^2_0}{4} \tr{a_{\mu} a^{\mu}},
\la{bos}
\ea
where
$$\widehat H(x) \equiv \left(\begin{array}{c}
H(x) - M_0  -
\frac{\delta f_0}{\Lambda^2}  a_\mu  a_\mu\\
\\
H^\dagger (x) - M_0  -
\frac{\delta f_0}{\Lambda^2}   a_\mu  a_\mu
\end{array} \right)$$
\ba
\widehat g^{-1} \equiv \left(\begin{array}{cc}
\frac{1}{g_{-}} & \frac{1}{g_{+}}
\\
\frac{1}{g_{+}} & \frac{1}{g_{-}}
\end{array}\right) ,
\ea
with
\be
\frac{1}{g_{-}}\equiv
\frac{1}{g_{S0}} - \frac{1}{g_{P1}},\qquad
\frac{1}{g_{+}}\equiv
\frac{1}{g_{S0}} + \frac{1}{g_{P1}}.
\ee

After integrating over fermion fields
one arrives to the following effective action for the
colorless bosonic variables
$\xi,  \Sigma ,\widetilde\Pi ,$
\be
 {\cal W}  = \int d^4x \,{\cal L}_{bos}  - N_c
{\rm Tr}\log\left(
\frac{ \not\! D +
  H(x)P_L + H^+(x) P_R}{\mu}\right)_{\Lambda} . \la{act}
\ee
Here the Tr operation stands for the total trace
in coordinates and matrix indices. Notice
that since $f_0^2$ is known to be proportional to $N_c$, the
whole effective action is proportional to $N_c$. In the large
$N_c$ limit the corresponding functional integral will be
saturated by its extrema.

Our starting wilsonian action had all modes above
the chiral symmetry breaking scale $\Lambda$ integrated
out. The information about higher frequencies was contained
in the coefficients of the intervening local operators.
Obviously, to avoid double counting, we must now
integrate the modes of the fermions only up to the scale
$\Lambda$. This is denoted by the subindex $\Lambda$
in the trace in \gl{act}. We shall in a few lines
discuss how we do so. Notice also that
$\Lambda$ is also the dimensional parameter normalizing
higher dimensional operators in the
ECQM action \gl{ECQM}.
The scale $\mu$ is an arbitrary
 normalization scale for quantum
contributions to the fermion free energy and
is also an essential ingredient in the preparation of
the low energy effective action (see below).

The Dirac operator appearing in \gl{act}, whose determinant
we must compute,
is of the form
\be
 \hat D \equiv i (\not\!\partial+ \not\! V  - \gamma_5\not\! A) +
 i ( S + i \gamma_5 P). \la{exts}
\ee
Given that we do not consider here anomalous,
parity odd processes involving pseudoscalar fields, we only
need the real part of the determinant, which is
just the square root of
the determinant of the positive operator
\ba
\hat D \hat D^\dagger &=& - \not\! D_V^2 +
\gamma_5 \{\not\!{D_V}, \not\!{A}\} + S^2 + P^2  - i\gamma_5 [S, P]\no
&& + \not\!{D_V} S - i \{\not\! A, P\}
- \gamma_5 \{\not\!{A}, S\} + i\gamma_5
[\not\! D_V, P].
\ea
where $D_V \equiv \not\!\partial+ \not\! V$. We can
now define our regulator
\be
 \mbox{Re}{\rm Tr}\,\log\left(\frac{\hat D}{\mu}\right)_{\Lambda} \equiv
 \frac12 {\rm Tr}\left(
\Theta\left(\Lambda^2 - \hat D \hat D^\dagger\right)
\log\left(\frac{\hat D \hat D^\dagger}{\mu^2}\right)\right),
\la{cut}
\ee
where $\Theta(x)$ is the step-function.

Note that the cyclic property of the trace operation allows
us to prove the invariance of \gl{cut} under local chiral
transformations  of quark fields and/or  external sources
\be
\hat D (V, A, S, P) =  \Omega \hat D (V^\Omega, A^\Omega,
S^\Omega, P^\Omega) \bar\Omega^\dagger,
\ee
\be
{\rm Tr}\,\log\left( \frac{
\Omega \hat D \hat D^\dagger
(V^\Omega, A^\Omega,
S^\Omega, P^\Omega)
\Omega^\dagger}{\mu^2}\right)_{\Lambda} =
{\rm Tr}\,\log\left(\frac{\hat D \hat D^\dagger
(V^\Omega, A^\Omega,
S^\Omega, P^\Omega)}{\mu^2}\right)_{\Lambda}
\ee
where
\ba
&&\Omega = P_L \Omega_L + P_R \Omega_R\quad
\bar\Omega = P_L \Omega_R + P_R \Omega_L,\no
&& L_{\mu}^\Omega \equiv V_\mu^\Omega + A_\mu^\Omega =
\Omega_L^\dagger  (V_\mu +
A_\mu)\Omega_L + \Omega_L^\dagger \partial_\mu \Omega_L ,\no
&&R_{\mu}^\Omega \equiv V_\mu^\Omega -
A_\mu^\Omega = \Omega_R^\dagger  (V_\mu -
A_\mu)\Omega_R + \Omega_R^\dagger \partial_\mu \Omega_R ,\no
&& S^\Omega + i P^\Omega = \Omega_R^\dagger(S + i P)
\Omega_L,\no
&&S^\Omega - i P^\Omega =
\Omega_L^\dagger(S - i P)\Omega_R .
\ea
Note that
$\Omega \bar\Omega^\dagger \not= 1$.
In particular, one can choose
the transformation \gl{chv} with
$\Omega_R = \xi,\quad \Omega_L =\xi^\dagger$ and then the Dirac
operator in \gl{bos} is reproduced when substituting in
\gl{exts}
\be
V_\mu = \frac12 v_\mu,\quad A_\mu = - \frac12
g_A a_\mu,\quad S + i P = H = \Sigma + i \widetilde\Pi.
\la{chim}
\ee

The determinant of $\hat{D}$ will be expanded in inverse
powers of the physical scale $\Lambda$. In addition to
$\Lambda$ we
have two more scales, namely the subtraction scale
$\mu$ and,
since chiral symmetry is broken, we also expect
$\langle S\rangle=\langle H\rangle\neq 0$. The scale
$\mu$ can appear only within logarithms and is normalized
either by the scale
$\langle S\rangle$ acting as an infrared cut-off, or by the scale
$\Lambda$ acting as an ultraviolet cut-off.
Within each order in the expansion in inverse powers
of $\Lambda$ we in turn expand in inverse powers
of $\langle S\rangle$.
It can be shown\footnote{Compare
with \ci{16} where a different regularization was used.}
that for $\Lambda \gg \, \langle S \rangle $
and for $\mu \gg \langle S\rangle$
the leading
contribution of this expansion is given by
\ba
&&  \int d^4x\,{\cal L}_{1-loop} \equiv
-\frac12 {\rm Tr}\left(
\Theta\left(\Lambda^2 - \hat D \hat D^\dagger\right)
\log\left(
\frac{\hat D \hat D^\dagger}{\mu^2}\right)
\right)
 \simeq \no
&&\frac{N_c}{16\pi^2} \int d^4x\,
\Bigl\{     2 \Lambda^4
\left(\ln\frac{\mu^2}{\Lambda ^2} + \frac12 \right) - 2 \Lambda^2
\left(\ln\frac{\mu^2}{\Lambda ^2}+1 \right) \tr{S^2 + P^2}
\no
&& +\ln\frac{\mu^2}{\langle S \rangle^2}\,
 {\rm tr}\, \Bigl(
 (S^2 + P^2)^2 - [S, P]^2 \no
&& + [D^V_\mu, S]^2 + [D^V_\mu, P]^2
- \{A_\mu, S\}^2 - \{A_\mu, P\}^2 \no
&&- 2i [D^V_\mu, P] \{A_\mu, S\} +
2i [D^V_\mu, S] \{A_\mu, P\} \no
&&- \frac13 \left( (F_{\mu\nu}^L)^2 + (F_{\mu\nu}^R)^2\right)\Bigr)
    \Bigr\} ,\la{log}
\ea
where we have kept, of all terms of ${\cal O}(\Lambda^0)$,
only those which are logarithmically enhanced.
We thus neglect terms of ${\cal O}(p^4)$ which are not
logarithmically enhanced and  terms of
${\cal O}(p^6/\langle S\rangle^2)$ or
${\cal O}(p^6/\Lambda^2)$ or smaller (neither of the
latter have any logs).
The universal part of the bosonized quark determinant is given
by the logarithmic terms,
whereas the coefficients
of next-to-leading terms do substantially depend on the shape of
the regulator. In the retained terms all the information concerning the
preparation of the effective action is contained in only two parameters:
$\Lambda$ and $\mu$. The approximation based on retaining the
action \gl{log} is provided by the large-$\Lambda$ and
large-log limit. Of course we could well have used another regulator
instead of the $\Theta$ function, such as e.g. an exponential
damping factor, to compute the fermion determinant. If so
the terms not included in \gl{log}, which are non-universal,
 would
change, but so would the bare coefficients of our effective
 action at scale $\Lambda$, since they are defined
by integrating out modes above $\Lambda$. The result should indeed
be scheme independent.

After substitution of \gl{chim} in \gl{log} one obtains the bosonic form
of the ECQM action \gl{bos} with the lagrangian:
\be
{\cal L}(\Sigma, \xi, \tilde\Pi) = {\cal L}_{bos} +
{\cal L}_{1-loop} .
\la{boslog}
\ee
The saddle point of the action \gl{act} obtained from the
above lagrangian  is
the relevant configuration for perturbing around
in the large $N_c$ limit.
Due to the chiral symmetry the effective action
does not depend on constant chiral fields $\xi_0$.
The effective potential  for constant $\langle H \rangle \equiv
H_0 = \Sigma_0 + i
\Pi^a_0 \tau^a$
can then  easily be derived
\ba
V_{eff}(\Sigma_0, \Pi^a_0, M_0)
&=& N_c \Biggl\{\frac{ \Lambda^2}{g_{S0}} (\Sigma_0 - M_0)^2 +
\frac{ \Lambda^2}{g_{P1}} (\Pi_0^a)^2
+ \frac{1}{8 \pi^2} \left[- 2 \Lambda^2 |H_0|^2
\ln\frac{\mu^2}{\Lambda ^2}
\right.\no
&&\left.
 + |H_0|^4\left( \ln\frac{\mu^2}{|H_0|^2} +\frac12\right)
 + \Lambda^4
\ln\frac{\mu^2}{\Lambda^2}  \right]\Biggr\}.
\la{pot}
\ea
where $|H_0|^2=\Sigma_0^2 +(\Pi_0^a)^2$.
In the above expression for the effective potential
we have already retained the universal
logarithmically enhanced terms
only. The finite constants (except one which
has been kept for later convenience) have been absorbed by a finite
redefinition of the appropriate coupling constants.
The saddle-point equations (or mass-gap equations)
will determine the extremum of
$V_{eff}$
\be
\frac{\partial V_{eff}}{ \partial \Sigma_0} = 0
\Longleftrightarrow
 \frac{ \Lambda^2}{g_{S0}}\left(\Sigma_0 - M_0\right)
= \frac{\Sigma_0}{4\pi^2} \left(
\Lambda^2\ln\frac{\mu^2}{\Lambda^2} -  |H_0|^2
\ln\frac{\mu^2}{|H_0|^2}
\right), \la{massg1}
\ee
and
\be
\frac{\partial V_{eff}}{ \partial\Pi_0^a} = 0
\Longleftrightarrow
 \Pi^a_0 \left[\frac{ \Lambda^2}{g_{P1}} - \frac{1}{4\pi^2}\left(
\Lambda^2 \ln\frac{\mu^2}{\Lambda ^2}  -  |H_0|^2
\ln\frac{\mu^2}{|H_0|^2}
\right)\right] = 0, \la{massg2}
\ee

Evidently, the above effective potential has to be renormalized
to eliminate the dependence on the arbitrary scale $\mu$.
To eliminate the $\mu$ dependence we redefine
the four fermion coupling constants and the chiral invariant
mass $M_0$
\be
\frac{1}{g_{S0}(\Lambda)}\equiv
\frac{1}{g_{S0}(\mu)} -\frac{1}{4\pi^2}
\ln\frac{\mu^2}{\Lambda^2},
\qquad
\frac{1}{g_{P1}(\Lambda)}\equiv \frac{1}{g_{P1}(\mu)} -
\frac{1}{4\pi^2}\ln\frac{\mu^2}{\Lambda ^2},\la{RG1}
\ee
\be
M_0(\Lambda)=M_0(\mu)\frac{g_{S0}(\Lambda)}{g_{S0}(\mu)},
\ee
or, in differential form,
\be
\mu\frac{\partial g_{S0}}{\partial\mu}= -\frac{g_{S0}^2}{2\pi^2},
\qquad
\mu\frac{\partial g_{P1}}{\partial\mu}= -\frac{g_{P1}^2}{2\pi^2},
\qquad
\mu\frac{\partial M_0}{\partial\mu}=-g_{S0}\frac{M_0}{2\pi^2}.
\ee
These renormalization group equations coincide with those
obtained by diagrammatic methods in the large $N_c$ limit.
They are not enough, however, to
fully remove the $\mu$ dependence. Some pieces not originally present
in the ECQM are needed to render $V_{eff}$ truly
$\mu$ independent.
They are related to quark self-interaction operators of
dimension higher than 6. In particular, the logarithmic terms proportional
to $| H |^4$ in \gl{pot} are related to eight-fermion operators of dimension
12 which generate similar $| H |^4$- terms after bosonization
(see \ci{7}). In the bosonic language this implies adding
a piece $\lambda(\mu)|H|^4$ in $V_{eff}$ where
\be
\mu\frac{\partial \lambda}{\partial\mu}=-\frac{1}{4\pi^2}.
\ee
That such higher dimensional operator (in the
original fermionic variables) is required
is due to the fact that chiral and canonical dimensions
do not coincide. In addition we need to add
the field independent pieces
\be
-\frac{N_c\Lambda^2 M_0^2(\mu)}{g_{S0}(\mu)}-
N_c\Lambda^4\ln{\frac{\mu^2}{\Lambda^2}}
\la{add}
\ee
to the vacuum energy. Then $dV_{eff}/d\mu=0$.

Note that the renormalization group does not
uniquely fix the form of the effective potential. We can in fact add to the
low-energy  effective potential $V_{eff}$
any dimension four
polynomial constructed from the renormalization group invariants
$\Lambda^2$ and $M_0/g_{S0}$ and still have $dV_{eff}/d\mu=0$.  These
pieces would correspond to bare contributions which are not logarithmically
enhanced.

Matching with QCD suggests that just below the  CSB
scale the coupling constants $g_{S0}$,$g_{P1}$,
$\lambda$, etc.
must be in some sense small (for $M_0$ this means $M_0\ll\Lambda$).
We also expect the total constituent mass to satisfy
$\langle H_0\rangle \ll \Lambda$. We would expect these constants
to grow as we move away from the scale $\Lambda$ and we depart
more and more from perturbative QCD. Indeed, the previous
renormalization group analysis shows that
these constants
run and, for instance in the case
of $g_{S0}$, $g_{P1}$,  and $M_0$ grow as the energy decreases, which
is just what we  naively expect.

After renormalizing all the coupling constants to the value they take at
the scale $\Lambda$, the effective potential in the
leading log approximation  takes the form
\ba
V_{eff}(\Sigma_0, \Pi^a_0, M_0)  &\simeq& V_{eff}( 0, 0, M_0) \no
&& +  N_c \Biggl\{\frac{ \Lambda^2}{g_{S0}} (\Sigma_0^2 - 2M_0\Sigma_0) +
\frac{ \Lambda^2}{g_{P1}} (\Pi_0^a)^2\no
&& + \frac{1}{8 \pi^2} |H_0|^4 \ln\frac{\Lambda^2}{|H_0|^2}
\Biggr\},
\la{potR}
\ea
where we have neglected the term proportional to
$\lambda$ as well as any remaining non-logarithmic pieces.
Unless explicitly stated otherwise, from now on we will take our coupling
constants to be defined at scale $\Lambda$; that is, at the
CSB scale.

We expect the pseudoscalar coupling $g_{P1}$ to be
bounded in its negative values
\be
\frac{ \Lambda^2}{g_{P1}}  > - \frac{|H_0|^2}{4\pi^2}
\ln\frac{\Lambda^2}{|H_0|^2}. \la{pbr}
\ee
Indeed if the opposite inequality held then
the dynamical breaking of the $SU(2)$ isospin symmetry down to a
$U(1)$ subgroup would occur and
$ \langle\widetilde\Pi^3\rangle \equiv \Pi_0^3  \not= 0$. As it also would
break  $P$ we cannot accept
to have this phase in the hadrodynamics at normal conditions.
If \gl{pbr} holds then the real scalar
v.e.v. $\langle H\rangle = \Sigma_0, \langle\widetilde\Pi^a\rangle= 0$
delivers the appropriate solution of the mass gap equations
and determines the minimum of the effective potential.

As for the effective scalar coupling it may take a priori any values.
Let us first consider the case where $g_{S0}\sim {\cal O}(1)$.
In terms of the effective coupling the mass-gap equation reads
\be
 \frac{ \Lambda^2}{g_{S0}}\left(\Sigma_0 - M_0\right)
= - \frac{\Sigma^3_0}{4\pi^2} \ln\frac{\Lambda^2}{\Sigma_0^2} . \la{mgs}
\ee
Just as we expect $M_0\ll\Lambda$,
we presume that the total constituent mass $\Sigma _0 \ll\Lambda$.
Solutions to the gap equation exist
for
$g_{S0} > 0$
\be
\Sigma _0 = M_0 \left(1 + O\left(\frac{M_0^2}{
\Lambda^2}\right)\right), \qquad \Sigma _0 < M_0. \la{sigma}
\ee
For $g_{S0} < 0$ this solution does not provide a minimum.
On the contrary solutions for negative $g_{S0}$ exist for $g_{S0}$
large enough, which is best studied by introducing
\be
\bar g_S \equiv
g_{S0}\frac{M_0^2}{\Lambda^2} I_0,\qquad
\bar g_P \equiv
g_{P1}\frac{M_0^2}{\Lambda^2} I_0,\qquad
I_0\equiv\frac{1}{4\pi^2}\ln\frac{\Lambda^2}{M_0^2}.
 \la{strong}
\ee
We will call from now on strong coupling regime the one characterized by
$\bar g_S\simeq \bar g_P \simeq {\cal O}(1)$ and, accordingly, the weak
coupling regime will correspond to $\bar g_S\simeq \bar g_P\ll 1$. The
effective potential written in these variables
reads
\ba
V_{eff}(\Sigma_0, \Pi^a_0, M_0) &\simeq & V_{eff}( 0, 0, M_0) \no
&& +  N_c I_0 \Biggl\{\frac{M_0^2}{\bar g_{S}} (\Sigma_0^2 - 2M_0\Sigma_0)
+ \frac{ M_0^2}{\bar g_{P}} (\Pi_0^a)^2
+ \frac{1}{2} |H_0|^4
 \Biggr\}.
\la{potRR}
\ea
It is also convenient to introduce the variable
$x =  \Sigma_0 / M_0 $. With its help the mass-gap equation,
retaining the logarithmically enhanced terms only, looks  as follows
\be
\bar g_S x^3 +  (x - 1) \simeq 0 . \la{dena}
\ee
This expression shows that $\bar g_S$ (and $\bar g_P$) are the
natural variables in terms of which we can speak of weak or
strong four fermion interactions. Notice that
the original variables $g_{S0}$ and $g_{P1}$ were
accompanied only by the $N_c$ and $\Lambda$ factors to make them
dimensionless and of ${\cal O}(1)$ in the large $N_c$ limit, but
it was not clear which was the natural normalization.
For positive $\bar g_S$ it is evident that the equation \gl{dena}
has only one real, positive solution $0 < x < 1$. Its asymptotic
behaviours are
\be
x \sim 1, \quad \bar g_S \ll 1, \qquad x \sim \left(\bar g_S \right)^{-
1/3}, \quad \bar g_S \gg 1,
\ee

As mentioned above, in the strong coupling regime one can reach
`overcritical' values of the scalar coupling, i.e. $\bar g_S < 0$.
In this case the mass-gap eq.\gl{dena} has always one negative solution
which corresponds to the absolute minimum of the effective potential.
The asymptotic expression for this solution is
\be
x \sim \left(\bar g_S \right)^{- 1/3},
\quad \bar g_S \ll -1;\qquad
x \sim - \left|\bar g_S \right|^{- 1/2},
\quad 0 > \bar g_S \gg -1,
\ee
Evidently the solution  $x < 0$
always exists whereas real positive solutions arise
only for $0 > \bar g_S \geq - 4/27$ and
they never provide comparable minima.
In all cases physically acceptable solutions
must lie in the range
\be
\frac{\Lambda}{|M_0|}\gg\vert x\vert\gg\frac{|M_0|}{\Lambda}.
\ee
We notice that the sign of the total
constituent mass $\Sigma_0$ and the sign (so far unspecified)
of $M_0$ are opposite in the
overcritical regime
in order to deliver an absolute minimum of the effective potential.

\bigskip

\noindent{\bf 5. Mass spectrum of scalar and pseudoscalar states}

\medskip

In order to find the mass spectrum of bosonic states in the large $N_c$
limit  we expand the effective action ${\cal W}$ in scalar,
$\tilde\sigma \equiv \Sigma - \Sigma_0$,  and
pseudoscalar, $\widetilde\pi, \, \widetilde\Pi$ fluctuations
around  the corresponding vacuum expectation values.

We can read the kinetic term for the scalar
meson from \gl{log}, \gl{pot}
\be
{\cal W}^{(2)}_S = \frac12\int \frac{d^4p}{(2\pi)^4}
 \tilde\sigma(p) K_S (p^2) \tilde\sigma(-p),
 \la{W2}
\ee
where
\be
 K_S (p^2) =
N_c \left[\frac{2\Lambda^2}{g_{S0}} +
\left(6 \Sigma_0^2 + p^2\right) I_0\right].
\la{W3}
\ee
The normalization of the kinetic term leads to the physical
scalar field $\sigma$
\be
\sigma = \sqrt{N_c I_0}\tilde\sigma. \la{phys1}
\ee
Its mass is
\be
m^2_\sigma =
\frac{2  \Lambda^2}{g_{S0} I_0} + 6 \Sigma_0^2 .
\la{scmass}
\ee
The kinetic term for pseudoscalar states is derived by means of
the expansion of the effective action  in powers of
$\widetilde\Pi^a$ and
$a_\mu \simeq i\partial_\mu \widetilde\pi(x) /F_0$. The calculation
of the quadratic part respecting the chiral symmetry can be done
directly from the action \gl{log}, the effective potential \gl{pot}
and the mass-gap equation
\gl{mgs}
\ba
{\cal W}^{(2)}_P &=& \frac12\int \frac{d^4p}{(2\pi)^4}
\left[ \widetilde\pi^a(p) \frac{p^2}{F_0^2}\left(f_0^2 +
N_c g_A^2 \Sigma_0^2 I_0\right) \widetilde\pi^a(-p)\right.\no
&&\left.-  \widetilde\pi^a(p)
 p^2\frac{2N_c g_A \Sigma_0 I_0}{F_0} \widetilde\Pi^a(-p)\right.\no
&&\left. +
\widetilde\Pi^a(p) N_c \left(p^2 I_0 + \frac{2\Lambda^2}{g_{P1}} +
2 \Sigma_0^2 I_0\right)\widetilde\Pi^a(-p)
\right] .
 \la{WP2}
\ea
This form can be diagonalized and normalized with the choice
of the bare pion decay constant
\be
F_0 = \sqrt{f^2_0 + N_c \Sigma_0^2 g^2_A I_0}, \la{fpi}
\ee
and thus one determines the physical fields $\pi, \Pi$
\be
\widetilde\pi^a =
 \pi^a
+ \frac{\Sigma_0 g_A \sqrt{N_c I_0}}{f_0} \Pi^a,\qquad
\widetilde\Pi^a = \frac{F_0}{f_0\sqrt{N_c I_0}} \Pi^a .
\la{physP}
\ee
As expected one obtains one massless and one massive pseudoscalar
state with a mass
\be
m^2_{\Pi} = \frac{2 F_0^2}{I_0 f_0^2} \left( \frac{\Lambda^2}{g_{P1}}
+  \Sigma_0^2 I_0 \right),
\la{psmass}
\ee
where the relation \gl{fpi} has been used.

In the weak coupling regime terms quadratic in $\Lambda$
dominate in \gl{scmass}, \gl{psmass}
\be
m^2_\sigma \simeq
\frac{2 \Lambda^2}{g_{S0}I_0};\qquad
m^2_{\Pi} \simeq \frac{2 F_0^2 \Lambda^2}{g_{P1} I_0 f_0^2} \equiv
m^2_\sigma  \frac{\gamma}{\delta^2} \quad \mbox{with}\quad
\gamma \equiv \frac{g_{S0}}{g_{P1}};\quad \delta \equiv
\frac{f_0}{F_0} < 1. \la{weak}
\ee
Both masses are generated at the scale $\Lambda$ and due to
$I_0$ in the denominator lighter than $\Lambda$ by a logarithmic factor.
Notice that if both masses are comparable, i.e. $m_\sigma^2/m_\Pi^2=
{\cal O}(1)$, then $f^2_0/F^2_0={\cal O}(1)$ and $f_0^2$ must
also be logarithmically enhanced.

In the strong coupling regime it is more natural to use  $\bar g_S$
and $\bar g_P$.
The corresponding spectrum is then given by eqs. \gl{scmass}, \gl{psmass}
and in terms of the notations in \gl{dena},  \gl{weak}
\be
m^2_\sigma = 2 M_0^2 (\frac{1}{\bar g_S} + 3 x^2),\qquad
m^2_{\Pi} =
\frac{ 2 M_0^2}{\delta^2} (\frac{\gamma}{\bar g_S} + x^2). \la{scps}
\ee
The ratio of the masses is controlled by the constants $\delta$ and
$\gamma$ for different values of $\bar g_S$. We notice that in terms of
the rescaled constants $ \bar g_S ,\, \bar g_P $ the major
mass scale is $M_0$ and the scale $\Lambda$ is involved only in the
definition of
the parameter $I_0$.

\bigskip

\noindent{\bf 6. External sources, $F_{\pi}$ and current quark masses}

\medskip

External vector, axial-vector, scalar and pseudoscalar sources are
included in the basic QCD quark action with the help
of the extension \gl{exts}
of the Dirac operator.  In the low energy effective action below
the chiral symmetry breaking scale implemented by the
ECQM, external sources are coupled to fermions in the same
way, but, in addition, they appear through  the
derivatives acting on chiral fields
\ba
&&\partial_\mu U \longrightarrow  D_\mu U \equiv
\partial_\mu U  + [\bar V_\mu, U]- \{\bar A_\mu, U\},\no
&& l_\mu = U^\dagger  D_\mu U,
\qquad r_\mu = -  D_\mu U  U^\dagger.
\ea
It is known that in this way the CVC and PCAC identities
hold.
As a consequence, in the rotated
basis of constituent quark fields, $Q_L,Q_R$, one arrives
to the following modification of the induced
effective vector and axial-vector fields
in \gl{chim}
\ba
&&V_\mu = \frac12 \left( \xi^\dagger \partial_\mu \xi -
\partial_\mu \xi  \xi^\dagger +  \xi^\dagger \bar V_\mu \xi +
\xi \bar V_\mu\xi^\dagger - \xi^\dagger \bar A_\mu \xi +
\xi \bar A_\mu \xi^\dagger\right)\no
&&A_\mu = -\frac12 g_A \left( \xi^\dagger \partial_\mu \xi +
\partial_\mu \xi \xi^\dagger +  \xi^\dagger \bar V_\mu \xi -
\xi \bar V_\mu\xi^\dagger - \xi^\dagger \bar A_\mu \xi -
\xi \bar A_\mu \xi^\dagger\right). \la{long}
\ea
When substituted in \gl{log} they provide
the bosonized
matrix elements
of quark currents by performing the appropriate functional
derivatives. In particular this allows us to determine
the weak decay constant of the pion
after an
axial-vector source $\bar A_\mu$ is incorporated into the lagrangian.
Linearizing the matrix $U$, this amounts to the replacement
\be
\partial_\mu\widetilde\pi \longrightarrow \partial_\mu\widetilde\pi
+ 2 i F_0 \bar A_\mu ,
\ee
which produces the following piece in the lagrangian
\be
- i F_0 \tr{\partial_\mu \pi \bar A_\mu} \la{fpi1}
\ee
and thus we conclude that $F_\pi=F_0$,
with $F_0$ given by \gl{fpi}.
One can also
check that the decoupling of the additional
heavy pseudoscalar field $\Pi$ from
the axial current is exact in the chiral limit and that
the chiral symmetry is realized in the EPCAC way\ci{17}.
The identification of $F_\pi$ is slightly more subtle
away from the chiral limit.

Let us introduce the current quark mass $m_q>0$
and, at the
same time, prepare our effective action for
the derivation of two-point correlators for scalar
and pseudoscalar currents. These correlators can be obtained
from the vacuum generating functional of QCD if one incorporates
scalar $\bar S(x)$ and pseudoscalar $\bar P(x)$
external sources by functional derivation with respect to them.
We supplement  \gl{ECQM} with
\be
\Delta{\cal L}_{ECQM}
= i \bar q \left(m_q + Z_m (\bar S + i \gamma_5 \bar P)  \right) q.
\la{DelL} \ee
The constant $Z_m$ is introduced in a chirally symmetric way
to provide the counterterms for the
renormalization of the scalar and pseudoscalar densities (see
section 7).
The current mass can obviously
be reinterpreted as a non-zero constant value
for the source $\widetilde S(x) =  m_q + Z_m \bar S(x)$.
Eq. \gl{DelL} breaks chiral invariance, but this formally holds
if we assume that the external sources rotate appropriately
under chiral transformations
\be
{\cal M}\equiv m_q + Z_m (\bar S + i \bar P),\qquad
 {\cal M} \rightarrow \Omega_R {\cal M} \Omega^\dagger_L,\quad
q_L \rightarrow  \Omega_L q_L,\quad
q_R \rightarrow  \Omega_R q_R . \la{chinv}
\ee

After the explicit inclusion of the Goldstone boson degrees
of freedom in the effective action below the chiral symmetry
breaking scale we parametrize the residual gluon
interactions in terms of local operators. In addition to those
already considered, there are,
away from the chiral limit, quite a few more
that one can write by including the matrix
${\cal M}$ and using the transformations
\gl{chinv}.
One such operator is the counterpart of the
term with the bare pion decay constant $f_0$, which, as we have
seen, plays a crucial role to make the field $\Pi$ a
propagating one. It is
\be
c_1 \tr{ {\cal M} U^\dagger + {\cal M}^\dagger U} .
\la{lsp}
\ee
There are several operators linear in the scalar
source of chiral dimension $\le 4$. One of them is
\be
c_2 \tr{(U^\dagger{\cal M} +{\cal M}^\dagger U) l_\mu l_\mu},
\ee
and it is in fact the only relevant one
since $
\tr{{\cal M}U^\dagger +{\cal M}^\dagger U}
\tr{l_\mu l_\mu}$ is, in fact, not independent for $SU(2)$
and $\tr{(U^\dagger{\cal M} +{\cal M}^\dagger U)D_\mu l_\mu}$
can be reduced with the help of the EOM.

There are several operators quadratic in the current
masses. The relevant one is
\be
c_3 \tr{{\cal M} U^\dagger{\cal M} U^\dagger +U{\cal M}^\dagger U{\cal
M}^\dagger }, \la{llsspp}\ee
since
other operators quadratic in $m_q$ are either
subleading in $N_c$ or $U$-independent
\ba
&&\tr{{\cal M}U^\dagger} \tr{{\cal M}U^\dagger}+
\tr{{\cal M}^\dagger U} \tr{{\cal M}^\dagger U}
,\no
&&\tr{{\cal M}U^\dagger} \tr{{\cal M}^\dagger U},\no
&&\tr{{\cal M}{\cal M}^\dagger}.
\ea
We shall not go beyond this order and, in fact, we shall
consider two-point correlators in the chiral limit only.

In the fermion sector there are several operators with the required
invariance properties and linear in ${\cal M}$.
The possible dimension 4 operators without derivatives are
\ba
&&\bar q_R{\cal M} q_L + \bar q_L {\cal M}^\dagger q_R,\no
&&\bar q_R U {\cal M}^\dagger U q_L +
\bar q_L U^\dagger {\cal M} U^\dagger q_R,\no
&&\bar q_R U q_L \tr{{\cal M} U^\dagger} +
\bar q_L U^\dagger q_R \tr{U {\cal M}^\dagger},\no
&&\bar q_R U q_L \tr{U {\cal M}^\dagger} +
\bar q_L U^\dagger q_R \tr{{\cal M} U^\dagger}.  \la{mst}
\ea
Those containing the additional trace will be of subleading order
in the large $N_c$ expansion
and thus not considered here. Operators containing twice the source
${\cal M}$ give eventually contributions which are down
by a power of $\Sigma_0^2/\Lambda^2$ and will be neglected
accordingly.

Higher-dimensional operators with derivatives
containing ${\cal M}$ can be
reduced with the
help of the EOM either to the operators \gl{mst} or to
a set of operators which contribute to
the next-to-leading order of
the large $N_c$ expansion and therefore can be neglected.

In the rotated quark basis, the only two relevant
operators are as follows
\ba
&&\bar Q_R \xi^\dagger {\cal M} \xi^\dagger Q_L
+ \bar Q_L \xi {\cal M}^\dagger \xi Q_R ,\no
&&\bar Q_R \xi {\cal M}^\dagger \xi Q_L
+  \bar Q_L \xi^\dagger {\cal M} \xi^\dagger  Q_R. \la{massv}
\ea
Their existence is due to the nonlinear realization of the chiral
symmetry on the fields $ \xi,\xi^\dagger$ \gl{nlr}
resulting  in
a hidden vector symmetry representation on products
\ba
&&\xi {\cal M}^\dagger \xi \longrightarrow
h_\xi \xi {\cal M}^\dagger \xi h^\dagger_\xi,\no
&& \xi^\dagger {\cal M} \xi^\dagger \longrightarrow
h_\xi \xi^\dagger {\cal M} \xi^\dagger h^\dagger_\xi.
\ea
In principle these terms have different coupling constants unless
a hidden left-right symmetry holds, $Q_L \leftrightarrow Q_R$.
In the QCD low energy effective action one would not a priori expect
to have this symmetry. However its appearance could be
revealed from the meson phenomenology (see section 8).

Thus we supplement the ECQM lagrangian with the following
piece
\ba
\Delta{\cal L}_{ECQM}  &=&
i (\frac12 + \epsilon) \left(\bar Q_R \xi^\dagger {\cal M}
\xi^\dagger Q_L + \bar Q_L \xi {\cal M}^\dagger \xi Q_R \right)\no
&& + i (\frac12 - \epsilon) \left( \bar Q_R \xi {\cal M}^\dagger \xi Q_L
+  \bar Q_L \xi^\dagger {\cal M} \xi^\dagger  Q_R\right)\no
&& + c_1 \tr{ {\cal M} U^\dagger + {\cal M}^\dagger U} \no
&& + c_2 \tr{(U^\dagger{\cal M} +{\cal M}^\dagger
U)l_\mu l_\mu}\no
&& + c_3 \tr{{\cal M} U^\dagger{\cal M} U^\dagger + U{\cal M}^\dagger U{\cal
M}^\dagger } , \la{massi}
\ea
where $\epsilon$ and $c_1,c_2,c_3$ are real coupling constants.
The above mentioned hypothetical left-right  symmetry would correspond
to $\epsilon = 0$.
We notice that the choice of the
couplings is
made so that the mass term corresponding to ${\cal M} = m_q$ in the lowest
order chiral interaction,  $U = 1$, coincides
with the QCD mass term for low energy quark fields.

After the introduction of auxiliary
scalar and pseudoscalar fields just as in \gl{HS}
we perform a shift of variables to simplify the structure of the
quark determinant as in \gl{shi}. Thus we define
\be
\widetilde\Sigma + i \widetilde\Pi\equiv
 \tilde\sigma + i \widetilde\Pi' +
\Sigma_0 - M_0  -
\left( (\frac12 + \epsilon) \xi^\dagger {\cal M} \xi^\dagger
 + (\frac12 - \epsilon) \xi {\cal M}^\dagger \xi\right)
-\frac{\delta f_0}{\Lambda^2}a_\mu a_\mu ,
\la{shift}
\ee
with $\tilde\sigma=\Sigma-\Sigma_0$.

The shift \gl{shift} allows us to move the dependence
on scalar sources to the purely bosonic part
of the effective action ${\cal L}_{bos}$, which is now
modified by the following terms (we keep only those
which will be relevant in the following discussion)
\ba
\Delta{\cal L}_{bos}
&=& \left( c_1 -
\frac{N_c \Lambda^2 (\Sigma_0 - M_0)}{2 g_{S0}}\right)
\tr{ {\cal M} U^\dagger +  {\cal M}^\dagger U}\no
&& + \left( c_2+\frac{N_c\delta f_0}{2g_{S0}} \right)
\tr{(U^\dagger{\cal M}  +  {\cal M}^\dagger U)l_\mu l_\mu}\no
&& - \frac{N_c \Lambda^2}{2 g_{S0}}\tilde\sigma \tr{{\cal M} U^\dagger +
{\cal M}^\dagger U}-\frac{N_c\delta f_0}{g_{S0}}
\tilde\sigma \tr{l_\mu l_\mu}\no
&&+ i \epsilon \frac{N_c  \Lambda^2}{ g_{P1}}
\tr{\widetilde\Pi' ( \xi^\dagger {\cal M} \xi^\dagger
- \xi {\cal M}^\dagger \xi )} \no
&&+\left(c_3+\frac{N_c\Lambda^2}{8
g_{S0}}-\frac{4\epsilon^2N_c\Lambda^2}{8g_{P1}}\right)\tr{{\cal
M}U^\dagger{\cal M}U^\dagger +{\cal M}^\dagger U {\cal M}^\dagger U}
 . \la{del}
\ea
We notice that the difference between the two kinds of mass terms appears
only in the pseudoscalar channel. We also notice that there are two
different contact terms contributing to the scalar-axial-axial
three point function, one proportional to $c_2$, the other proportional
to $\delta f_0$. In fact, as it will be discussed in the coming
section in more detail, QCD dictates that the sum of the two pieces
just above the scale $\Lambda$ must be zero. The only sizeable
contribution to the scalar-axial-axial vertex is the one
induced by the exchange of the $\sigma$ particle, which indeed
vanishes as $1/p^2$. Finally, the scalar and pseudoscalar correlators
have also some contact terms, which can be read from the last line
in \gl{del}. QCD has also something to say on that particular
combination of coefficients.

Let us analyze the influence of current masses of light quarks on the
mass-gap solution and on the effective potential. We neglect the
isospin breaking effects (with respect to $SU(2)_V$) and take
${\cal M} = m_q \mbox{\bf I}$. From \gl{del}
we immediately read the modification in part of the effective potential
linear in $m_q$ (taking into account that $\langle\tilde \sigma\rangle=
0$)
\be
\Delta V_{eff} = \left( 4 c_1 -
\frac{2N_c \Lambda^2 (\Sigma_0 - M_0)}{ g_{S0}}\right) m^U_q ,
\la{supl}
\ee
where
\be
m^U_q\equiv m_q \cos\phi,\quad
\cos\phi \equiv \frac14 \tr{U + U^\dagger}.
\ee
We shall assume, as is commonly done,
that $\langle \cos\phi\rangle =1$. Other choices
of the chiral vacuum lead to different
phases for $M_0$, the chiral condensate, etc.

Just as the other parameters in the theory, we have to renormalize
$m_q$ to absorb the logs associated to the
running between the scales $\mu$ and $\Lambda$. Its
renormalization can, for instance, be obtained diagrammatically
and is
identical to that of $M_0$
\be
\frac{m_q (\mu)}{g_{S0}(\mu)} =
\frac{m_q (\Lambda)}{g_{S0}(\Lambda)},
\qquad m_q (\Lambda) =
\frac{m_q (\mu)}{1 -\frac{g_{S0}(\mu)}{4\pi^2}
\ln\frac{\mu^2}{\Lambda^2}}, \la{RGm}
\ee
Another way of seeing this is by realizing that
to make the effective potential renormalization
group invariant in the presence of the soft breaking induced by the mass
term we have to add a piece similar to the first term in \gl{add}, namely
\be
-\frac{N_c\Lambda^2}{g_{S0}}(M_0+m_q^U)^2
\la{add1}
\ee
which, unlike \gl{add}, is now field dependent. Its origin
is obviously nothing but the terms \gl{lsp} and  \gl{llsspp}.
We thus see that it is necessary
to include the operator \gl{lsp} from the
beginning and that
\be
c_1(\Lambda)= -\frac{N_c\Lambda^2 M_0}{2 g_{S0}} + \tilde c_1(\Lambda),
\ee
where $\tilde c_1(\Lambda)$ contains other contributions which
cannot be deduced from renormalization group arguments alone (and
which for this reason are not expected to be logarithmically
enhanced either). In any case the combination
\be
2 c_1(\mu)  g_{S0}(\mu) + N_c \Lambda^2 M_0(\mu)=
2 c_1(\Lambda)  g_{S0}(\Lambda) + N_c \Lambda^2 M_0(\Lambda)
\la{RGc}
\ee
is a renormalization
group invariant.

The modified mass-gap equation is (from now on we take
$\langle \cos \phi\rangle =1$ and thus $m_q^U =m_q$)
\be
\frac{ \Lambda^2}{g_{S0}}\left(\Sigma_0 - M_0 -
m_q\right)
= - \frac{\Sigma^3_0}{4\pi^2} \ln\frac{\Lambda^2}{\Sigma_0^2}. \la{msg2}
\ee
Let us use the fact that
$m_q$ is a small parameter and let us calculate the solution
as a perturbation from the chirally symmetric one
\be
\Sigma_0(m_q) = \Sigma_0(0) + m_q \Delta
+O(m_q^2)
\qquad \Delta=
\frac{\Sigma_0(0)}{3 M_0 - 2 \Sigma_0(0)} < 1  . \la{Sig1}
\ee
In the weak coupling regime the solution is
$\Sigma_0 \simeq M_0 + m_q$
with $ M_0 \gg m_q$ by assumption. In the strong coupling regime
it is helpful again to invoke the notation of \gl{strong} to present
the mass-gap equation in the form of \gl{dena}
\be
\bar g_S x^3 +  (x - 1 -\frac{m_q}{M_0}) \simeq 0, \la{dena1}
\ee
with the approximate solution
\be
x = x_0 + \frac{x_0}{3 - 2 x_0}\cdot\frac{m_q}{M_0} =
 x_0 + \frac{1}{3 \bar g_S x_0^2 +1 }\cdot\frac{m_q}{M_0},\qquad
x_0 \equiv \frac{\Sigma_0(0)}{M_0}. \la{masscor}
\ee
This expression is valid for both subcritical, $\bar g_S > 0$,
and overcritical, $\bar g_S < 0$, regimes.
As we shall see
later on (see \gl{L5}) the correct sign for
the chiral constant $L_5$ for all values of $g_S$
is obtained only if the bare constituent mass $M_0$
has the same sign as $m_q$ (positive in our conventions).
Due to the mass-gap
equation the total
constituent mass $\Sigma_0$ which corresponds to the absolute minimum
for the effective potential takes the opposite sign in the
overcritical
strong coupling regime. Thus $\Sigma_0$ is positive in the
weak coupling regime and in the subcritical strong coupling one,
but negative for overcritical values of $\bar g_S$. (See
the discussion at the end of the previous section.)

Now we derive
the correction to the value of the effective potential \gl{pot}
at the minimum due to the presence of
the current quark mass \gl{supl}. Substituting \gl{msg2}
into the effective potential (including \gl{supl}) and
keeping terms linear in $m_q$,
one finds for the vacuum energy
\be
V_{eff}^{min} =  V_{eff}^{min}(m_q = 0) +
 m_q \left(4 c_1  + \frac{N_c}{2\pi^2} \Sigma_0^3
\ln\frac{\Lambda^2}{\Sigma_0^2}\right).  \la{Vmin}
\ee
The physical quark
condensate in Minkowski space,
renormalized at scale $\Lambda$ will
be
\be
C_q=i\langle\bar q q\rangle_{eucl}=\frac{1}{2} \frac{\partial
V_{eff}^{min}}{\partial m_q}. \la{chircon}
\ee
Namely
\be
C_q (\Lambda)= \left(2 c_1
+ \frac{N_c}{4 \pi^2}
\Sigma^3_0
\ln\frac{\Lambda^2}{\Sigma_0^2}\right) . \la{cond1}
\ee

Expanding the chiral field $U$ around the identity
and using \gl{supl}
one finds the
Gell-Mann-Oakes-Renner formula \ci{13}
for the pion mass: $m^2_\pi F^2_0 \simeq - 2m_q C_q$.
 The pion mass has the correct physical
sign provided that
$C_q\le 0$. Since we will see later that
in the weak coupling regime $M_0$ and, consequently,
$\Sigma_0$ must be positive, the second term in the first line of
\gl{cond1} is positive and
we conclude that
$c_1$ must be negative and large enough, lest the pion
be a tachyon. Thus
 the bare coefficient $c_1$
provides a contribution to the condensate which is
bigger than the one induced by the fermion loop after
integration. Taking a look at \gl{add1} we will
see that in the large log approximation this is the case. In the
overcritical regime, $g_{S0} < 0$,
the constant $c_1$ is less crucial as the accompanying
 term becomes
negative.

Let us now calculate the effects due to the soft CSB induced
by the current quark mass. They arise from the
perturbed solution $\Sigma_0 (m_q)$ of the mass gap equation
\gl{Sig1} in the first place, but they also
modify
the pseudoscalar mass matrix from \gl{del}
\be
\Delta {\cal L}^{(2)}_P = \frac12 \left( 2m_q B_0 (\widetilde\pi^a)^2
+  \frac{8 N_c \Lambda^2}{g_{P1} F_0} \epsilon m_q
\widetilde\pi^a \widetilde\Pi'^a\right), \la{pscor}
\ee
with $B_0 \equiv |C_q|/F_0^2$.
We notice that the parameter $\epsilon$ cannot be in general
renormalization independent if $g_{P1}\not= g_{S0}$.
Its renormalization should be given by
\be
\epsilon(\mu) \frac{g_{S0}(\mu)}{g_{P1}(\mu)}  =
\epsilon(\Lambda)\frac{g_{S0}(\Lambda)}{g_{P1}(\Lambda)} .
\ee

After the diagonalization of the
kinetic term with the help of variables \gl{physP}
(with $\widetilde \Pi$ replaced by $\widetilde \Pi^\prime$)
the mass matrix in the pseudoscalar sector ${\bf M}_P$
is still non-diagonal away from the
chiral limit
\be
{\bf M}_P =\left(\begin{array}{cc}
m^2_\pi & d_1 m^2_\pi\\
 d_1 m^2_\pi & m^2_\Pi(0) + d_2  m^2_\pi
\end{array}\right), \la{massm}
\ee
where the following notations are employed
\ba
&&m^2_\pi = 2 m_q B_0,\no
&&d_1 =  \frac{2\sqrt{N_c} \Lambda^2\epsilon}{g_{P1} B_0 f_0 \sqrt{I_0}}
  + \frac{\sqrt{F_0^2 - f_0^2} }{f_0},\no
&& d_2 = \frac{\Delta}{B_0\Sigma_0} \left( m^2_\Pi(0)
\left(1 -  \frac{f_0^2}{F_0^2}\right) +
\frac{2F_0^2 \Sigma_0^2}{f_0^2}\right)
+ \frac{4 N_c \Lambda^2\Sigma_0 g_A \epsilon}{g_{P1} B_0 f_0^2}
 + \frac{N_c\Sigma_0^2 g_A^2 I_0 }{f_0^2}.
\la{denps}
\ea
Then, rediagonalization of
 \gl{massm} induces the coupling of
the physical heavy $\Pi$ meson to the axial current from \gl{fpi1},
something which is forbidden in the chiral limit. At
first order in the current quark mass the corresponding decay constant is
given by
\be
F_{\Pi} = F_0 d_1
\frac{ m^2_\pi}{m^2_\Pi(0)}. \la{fPi}
\ee
The mass corrections to $F_\pi$ will be discussed at the end of
next section.

For  completeness we also show
the mass shift for heavy scalar and pseudoscalar mesons
\ba
&&m^2_\sigma (m_q) = m^2_\sigma (0) +  m^2_\pi \frac{6 \Sigma_0 \Delta}{B_0}
 +  O(m^4_\pi),\no
&& m^2_\Pi(m_q) = m^2_\Pi(0) + d_2  m^2_\pi +  O(m^4_\pi).
\ea
For $SU(2)$ these corrections are very small.

\bigskip

\noindent{\bf 7. Correlators and chiral symmetry restoration}

\medskip

In large $N_c$ QCD the
two-point correlators of scalar and pseudoscalar
quark densities are given by a regulated sum
over the appropriate meson poles
\ba
\Pi_S (p^2) &=&- \int d^4x \,\exp(ipx)\
\langle T\left(\bar q q (x) \,\, \bar q q
(0)\right)\rangle \no
&=&
\sum_n \,\frac{Z^S_n}{p^2 + m^2_{S,n}} + C^S_0 + C^S_1 p^2,\no
\Pi_P^{ab} (p^2)&=& \int d^4x \,\exp(ipx)\
\langle T\left(\bar q\gamma_5 \tau^a q (x) \,\,
\bar q\gamma_5 \tau^b q (0)\right)\rangle \no
 &=&  \delta^{ab}\left(
\sum_n \,\frac{Z^P_n}{p^2 + m^2_{P,n}}  + C^P_0 + C^P_1 p^2\right).
\la{planar}
v\ea
$C_0$ and $C_1$ are regularization-dependent contact terms.

The high-energy asymptotics is controlled by
perturbation theory and the operator product expansion due
to the asymptotic freedom of QCD.
In QCD the conventional derivative,
$i\partial_{\mu}\rightarrow iD_{\mu}
=i\partial_{\mu} + G_{\mu}$, contains gluon fields
$G_{\mu} \equiv  g_s \lambda^a G^a_{\mu}$,
where $tr(\lambda^a \lambda^b) = 2 \delta ^{ab}$.
 In the chiral limit ($m_q = 0$)
the scalar and pseudoscalar
 correlators have the following
power-like behavior \ci{11,12} at large $p^2$
\ba
&&\Pi_{S,P}(p^2) |_{p^2 \rightarrow \infty} 
\,\simeq \, \frac{N_c}{4\pi^2}\left(
1 + \frac{11 (N_c^2 - 1)\alpha_s}{8  N_c \pi}\right) p^2\,\ln\frac{p^2}{\mu^2}
\no
&&+ \frac{\alpha_s}{4\pi p^2} \langle (G^a_{\mu\nu})^2\rangle
+ \frac{2\pi \alpha_s}{3 p^4}
 \langle \bar q \gamma_{\mu} \lambda^k q
\bar q \gamma_{\mu} \lambda^k q\rangle \no
&&
\mp \frac{\pi\alpha_s}{p^4} \langle \bar q \sigma_{\mu\nu} \lambda^k q
\bar q \sigma_{\mu\nu} \lambda^k q\rangle +
{\cal O}(\frac{1}{p^6})\la{as}
\ea
for $N_f = 2$, in euclidean notations. In the planar (large-$N_c$)
limit\footnote{
We should mention some discrepancies in the literature \ci{11,12,12a}
concerning
the signs of the last term in \gl{as} and
acknowledge S. Peris for bringing to our attention the paper \ci{12a} 
where the signs are proven to be correct. The correctness of the
above signs can be verified by realizing that they provide the
correct Weingarten inequality\ci{Wein}
between pseudoscalar and scalar two-point correlators when
applied to the leading (logarithmic) singularity at short distances.}
\ba
&&\left(\Pi_P(p^2)- \Pi_S(p^2)\right)_{p^2 \rightarrow \infty} \equiv
\frac{\Delta_{SP}}{p^4}  + 
{\cal O} \left(\frac{1}{p^6}\right);\no
&&\Delta_{SP} \simeq  \frac{24(N_c^2 - 1)}{N_c^2} \pi\alpha_s C_q^2
\simeq 24 B_0^2 F_0^4,   \la{CSR}
\ea
where the vacuum dominance hypothesis\ci{11}  has been applied and
we have taken\ci{10} for simplicity
$\alpha_s (1.2 \mbox{\rm GeV}) \simeq 1/3$.

This rapidly decreasing asymptotics is of course a consequence of the manifest
chiral invariance of massless QCD.
When comparing \gl{planar} and \gl{as} one concludes that
an infinite series of
resonances with the same quantum numbers should exist in every channel
in order to reproduce
the perturbative asymptotics.

At scales below $\Lambda$, where chiral symmetry is
dynamically broken, the previous equality needs not to be valid.
However, the fact that chiral symmetry is restored so quickly
at high energies suggests itself as a further constraint
to impose on the ECQM. We shall demand that at the scale $\Lambda$,
and within the precision of our approach,
the relation \gl{CSR} is approximately fulfilled. This we will call
the chiral symmetry restoration (CSR) constraint \ci{19} (similar CSR 
constraints in the vector-axial vector channels  for specific models
have been considered in \ci{DG,Mous}) . In this
section we shall work out its consequences in the chiral limit $(m_q=0)$ and
we will restrict ourselves mostly to the scalar
and pseudoscalar two-point functions, although some three point functions
will also be discussed.

In the ECQM the
 two-point
correlators in the scalar and pseudoscalar channels are obtained
by performing functional derivatives of the vacuum generating functional
${\cal Z}(\bar S, \bar P)$ of the ECQM
\ba
\Pi_S (p^2) &=& \int d^4x \,\exp(ipx) \frac{\delta}{\delta\bar S(x)}
\frac{\delta}{\delta\bar S(0)}\,\log {\cal Z} (\bar S, \bar P)|_{\bar
S,\bar P = 0}\no &=&\frac{Z_\sigma}{p^2 + m^2_\sigma} + C^S(p^2),\no
\Pi_P^{ab} (p^2) &=&
\int d^4x \,\exp(ipx) \frac{\delta}{\delta \bar P^a(x)}
\frac{\delta}{\delta\bar P^b(0)}\,\log {\cal Z} (\bar S, \bar P)|_{\bar
S,\bar P = 0}\no &=& \delta^{ab}\left(
\frac{Z_\pi}{p^2 } + \frac{Z_\Pi}{p^2 + m^2_\Pi}
  + C^P(p^2)\right),
\la{plan}
\ea
where in the chiral limit the masses $m_i$
are given by eqs. \gl{scmass},
\gl{psmass} and the parameters $Z_i$ will be derived below.
Since we are well below the heavier poles (located above $\Lambda$),
 the functions
$C^S(p^2), C^P(p^2)$ will be, by construction,
finite polynomials
in momenta in the ECQM
\be
C^S(p^2)=C^S_0+p^2 C^S_1+\ldots \qquad
C^P(p^2)=C^P_0+p^2 C^P_1+\ldots \qquad
\ee
With this structure in mind
the CSR constraint will basically involve the
composite states $\pi, \Pi, \sigma$
of the model and decouple from the effects
of heavier degrees of freedom. This decoupling
is always possible with a suitable choice of the coupling constants
of local operators quadratic in the external sources.

When imposed on
two-point correlators of the ECQM (in the chiral limit)
the CSR condition leads to
\ba
&& C_0^S = C_0^P,  \la{srul1}\\
&& Z_\sigma = Z_\pi + Z_\Pi, \qquad   Z_\sigma > Z_\Pi,\la{srul2}\\
&& Z_\sigma  m^2_\sigma =  Z_\Pi  m^2_\Pi + \Delta_{SP}. \la{srul3}
\ea
As for the constants $C_1^S$ and  $C_1^P$, the CSR constraint
is automatically fulfilled in the
minimal ECQM after imposing our selection rules since there
are no contact terms with two derivatives acting on ${\cal M}$ of chiral
dimension $\le 4$.
There is one  operator of chiral dimension 4 that, on dimensional grounds,
could contribute to the first condition \gl{srul1}, namely $\tr{{\cal M}
{\cal M}^\dagger}$, but it does not since scalar and pseudoscalar sources
appear on equal footing in it. However, \gl{srul1} does provide
some information on the ECQM, namely
\be
c_3+\frac{N_c\Lambda^2}{8
g_{S0}}-\frac{4\epsilon^2N_c\Lambda^2}{8g_{P1}}=0. \la{canc3}
\ee
The constraints \gl{srul2} and \gl{srul3} allow to
determine two  constants $Z_i$
in terms of  a third one and meson masses. In particular
\be
 Z_\sigma =  \frac{Z_\pi m^2_\Pi - \Delta_{SP}}{m^2_\Pi - m^2_\sigma},
\qquad
Z_\Pi =  \frac{Z_\pi m^2_\sigma - \Delta_{SP}}{m^2_\Pi - m^2_\sigma}.
\la{srsol}
\ee
PCAC dictates that
$Z_\pi = 4 B_0^2 F_0^2$. We can use this relation
in \gl{srsol}
to find the other $Z$'s. From phenomenology \ci{10} $m_\Pi \simeq 1.3$ GeV
and therefore $Z_\pi m^2_\Pi > \Delta_{SP}$ when taking into account 
\gl{CSR}. Inasmuch as $Z_\sigma$ is positive one derives from \gl{srsol} that
$m^2_\Pi > m^2_\sigma$.

In order to use the relations\gl{srsol} we must find the dependence
of the constants $Z_i$ on parameters of the ECQM.
For this purpose
we expand the appropriate part of the ECQM lagrangian ---extended
with scalar and pseudoscalar sources---
in powers of meson fields
\ba
\Delta{\cal L}_{ECQM}
&=&  Z_m\left[- \frac{2N_c  \Lambda^2}{g_{S0}}\tilde\sigma \bar S
-  \frac{4N_c  \Lambda^2 \epsilon}{g_{P1}} \widetilde\Pi^a \bar P^a\right.
\no &&\left.+
\frac{4}{F_0} \left( c_1 -
\frac{N_c  \Lambda^2 (\Sigma_0 - M_0)}{2 g_{S0}}\right)
\widetilde\pi^a \bar P^a\right] ,
\la{Deltal}
\ea
where $Z_m$ was introduced in \gl{DelL}.

It is evident that the constant $Z_m$ is required to renormalize
the operators in \gl{Deltal} and obeys the following
renormalization group equation
\be
\frac{Z_m(\mu)}{g_{S0}(\mu)} =
 \frac{Z_m(\Lambda)}{g_{S0}(\Lambda)}.
\ee
We shall take $Z_m (\Lambda)=1$. This amounts to fixing the normalization
of the external sources at a given scale and we do not lose any generality
by doing so. This convention simplifies the relation between $Z_\pi$ and the
quark condensate. Let us now diagonalize the kinetic term with the physical
fields \gl{phys1}, \gl{physP} and  adopt the following notations
\ba
&&\alpha \equiv  \frac{2N_c\Lambda^2}{g_{S0} \sqrt{N_c I_0}} =
 \frac{2\sqrt{N_c I_0} M_0^2}{\bar g_S},\no
&&\beta  \equiv \frac{4}{F_0\alpha }\left(
\frac{N_c \Lambda^2(\Sigma_0 - M_0)}{2 g_{S0}}  - c_1\right)
=  - \frac{2C_q}{F_0 \alpha} = \frac{B_0 F_0 \bar g_S}{\sqrt{N_c I_0} M_0^2}.
\la{denot}
\ea
Then the operators
 linear in fields and external sources relevant for the
derivation of two-point correlators can be written as
\ba
\Delta{\cal L}_{ECQM} &=&
- \alpha \sigma \bar S
- \alpha \beta \pi^a \bar P^a
- \frac{\alpha \left(2 \epsilon \gamma + \beta \sqrt{1-
\delta^2}\right)}{\delta} \Pi^a \bar P^a\no
&& =
\mp \sqrt{Z_\sigma}\, \sigma \bar S - \sqrt{Z_\pi }\,\pi^a \bar P^a \mp
\sqrt{Z_\Pi }\,\Pi^a \bar P^a. \la{Llin}
\ea
Notice that only the sign of the second term is fixed. The
sign of the other two terms depends on that of $\bar g_S$.
 From the first relation in \gl{srsol}
and comparing with \gl{Llin} one finds that
\be
\beta = \mbox{sign}(\bar g_S) 
\frac{\sqrt{1 - \frac{m^2_\sigma }{m^2_\Pi}}}{\sqrt{1 - 
\frac{\Delta_{SP}}{Z_\pi m^2_\Pi}}}
 ,\qquad \frac{\Delta_{SP}}{Z_\pi m^2_\Pi} \simeq \frac{6 F_0^2}{m^2_\Pi}
\simeq 0.03. \la{cons2}
\ee
Thus, from phenomenology \ci{10} $|\beta| < 1$.
The relation \gl{srul2} gives
\be
2\epsilon \gamma  = - \beta\sqrt{1 - \delta^2} \pm  \delta \sqrt{1 -
\beta^2},
\la{beta}
\ee
and, since $|\delta|< 1$, it follows that $|2\epsilon\gamma|\le 1$.

We can also obtain $F_\sigma$ from \gl{Llin}. The scalar decay constant is
defined through the relation (recall the equivalent one for $F_\pi$)
\be
2 B_0 F_\sigma =\sqrt{Z_\sigma}.
\ee
Therefore
\be
F_\sigma= \frac{\sqrt{N_c I_0}M_0^2}{B_0 \bar{g}_S}=\frac{F_\pi}{\beta}.
\la{Fsig}
\ee

We see that the CSR constraints represent a powerful tool
to reduce the number of input parameters of the ECQM.

Another phenomenological condition enabling us to restrict the parameters
of the ECQM is provided by the experimental determination of the chiral
constant $L_8$, which parametrizes the following operator in the low-energy
chiral lagrangian (in euclidean space)
\be
- 4 L_8 \, B_0^2\,\tr{{\cal M}^{\dagger}U
{\cal M}^{\dagger}U +
U^{\dagger}{\cal M} U^{\dagger}{\cal M}}.
\ee
For $U={\bf I}$ the above operator reduces to
\be
- 16  L_8 \, B_0^2\, \left( \bar S^2 - (\bar P^a)^2\right)
\ee
In the ECQM, after imposing the condition \gl{canc3}
the only
origin of this term is the exchange of
heavy meson states. This
leads precisely to the sum rule \gl{L8} displayed in section 2.
When combining it with the solution \gl{srsol}
of the CSR constraints \gl{srul2}, \gl{srul3}
and the definition \gl{Llin} for $Z_\pi$ one finds the remarkable
relation
\be
L_8 = \frac{1}{64 B_0^2} \left(\frac{Z_\sigma}{m^2_\sigma}
- \frac{Z_\Pi}{m^2_\Pi}\right) =
\frac{F_0^2}{16}\left(\frac{1}{m^2_\sigma} + \frac{1}{m^2_\Pi}\right)
\left(1 - \frac{\Delta_{SP}}{Z_\pi (m^2_\sigma + m^2_\Pi)}\right),
\la{L-8}
\ee
which represents one of the predictions of the ECQM and modifies
substantially  the estimates of \ci{19a}.
The term proportional to $\Delta_{SP}$ is very small and, in practice,
negligible
\be
\frac{\Delta_{SP}}{Z_\pi (m^2_\sigma + m^2_\Pi)} \simeq
\frac{6 F_0^2}{m^2_\sigma + m^2_\Pi} < 0.03. 
 \la{L-888}
\ee
Thus one expects that
\be
L_8 > \frac{F_0^2}{8 m^2_\Pi}.
\ee
This
is a consequence
of the CSR inequality \gl{srul3}, $m^2_\sigma < m^2_\Pi$, and
the smallness of $\Delta_{SP}$.

We finally turn to the chiral constant $L_5$, which
goes with the operator
\be
2\,B_0 \tr{(D_{\mu}U)^{\dagger}D_{\mu}U({\cal M}^{\dagger}U
+ U^{\dagger}{\cal M})}. \la{L555}
\ee
First of all, a contribution to this low energy
constant seems to come from
the operator proportional to the coefficient
$\left( c_2+\frac{N_c\delta
f_0}{2g_{S0}}\right)$
in the lagrangian \gl{del}. For $U = \mbox{\bf I}$ this operator
reduces to the term
\be
\tr{\bar S \bar A_\mu \bar A_\mu}, \la{contact}
\ee
which jeopardizes the correct chiral
high energy behavior of the following three-point correlator
\be
\Pi_{SAA} = \int d^4x d^4y \exp(ipx + i qy)
\left\langle T\left( \bar q q(0)\bar q \gamma_5 \gamma_\mu q(x)
\bar q\gamma_5 \gamma_\mu q(y)\right)\right\rangle. \la{saa}
\ee
For large values of momenta one expects from the
asymptotic freedom of QCD that
perturbation theory describes the leading asymptotics. But in
perturbation theory it is evident
that the correlator \gl{saa} identically vanishes
in the chiral limit as any Feynman diagrams necessarily
contain an odd number of
$\gamma$ matrices. It is a consequence of the chiral invariance under
the transformation
\be
q(x) \rightarrow \gamma_5 q(x),\qquad \bar q(x) \rightarrow
- \bar q(x)\gamma_5, \qquad \bar S(x), \bar P(x) \rightarrow - \bar S(x),
- \bar P(x), \ee
which in the chiral limit holds for the generating
functional of perturbative QCD.

Therefore the correlator \gl{saa} should vanish at large
momenta and cannot have a constant asymptotic behaviour
such as the one provided
by the contact term \gl{contact} in the generating functional.
We assume this behavior to hold at scale $\Lambda$
for the correlators of the ECQM and
impose one more CSR constraint
\be
 c_2 + \frac{N_c\delta f_0}{2g_{S0}}=0.
\ee
Consequently, the constant $L_5$ is induced solely
by
$\tilde\sigma$ exchange, both from  the terms in \gl{del} and from the piece
linear in $\tilde\sigma$ in \gl{log}
\be
-\frac{1}{2}N_c g_A^2\Sigma_0 I_0 \tilde\sigma \tr{l_\mu l_\mu}.
\la{extra}
\ee
By coupling an external axial
source and repeating the analysis done in the massless case, we determine
the shift in
the pion decay constant
\be
F^2_\pi = F_0^2 \left(1 + m^2_\pi \frac{\Delta}{B_0\Sigma_0}
\left(1 -  \frac{f_0^2}{F_0^2} + \frac{2N_c\Sigma_0\delta f_0}{F_0^2
g_{S0}}\right)\right)+
O(m^4_\pi),\la{fpic}
\ee
in the notations of \gl{Sig1}.
This implies
\be
L_5 = \frac{\Delta}{8 B_0\Sigma_0}(F_0^2 - f_0^2 +  \frac{2N_c\Sigma_0\delta
f_0}{g_{S0}}).\la{L5}
\ee
Notice the appearance of $\delta f_0$ in these formulae. As we announced,
this effective constant is relevant in the $m_q\neq 0$ case. This is a
nuisance because it introduces a further unknown parameter into the problem.
However,
it is clear from \gl{fpi} that $F_0^2 -f_0^2$
is logarithmically enhanced, something which is
not expected
to happen for $\delta f_0$. In the large log approximation it is thus
justified to neglect $\delta f_0$ in front of the other quantities
and we will
do so.

Notice also that \gl{L555} is linear in $m_q$
and the sign of $L_5$  is correlated with that of $\Delta$
and $\Sigma_0$. Indeed,  for any value of $\bar g_S$, the
product $\bar g_S x$ is positive, while the sign of $\Delta$ is
always the
same as the one of $x$. Thus the sign of $L_5$ is finally that of
$M_0$ and
we conclude that $M_0$ must be positive, just like $m_q$. Had we
chosen the
opposite sign for $m_q$ the sign of $M_0$ would have came out
reversed as well.

\bigskip

\noindent{\bf 8. Fit and discussion}

\medskip

Let us enumerate all the parameters of the ECQM. The first
seven parameters are not directly observable:
$ M_0$, $x$ (instead of $\Sigma_0$),
$\delta$ (instead of $f_0$), $\bar g_S$, $\gamma$ (instead of
$g_P$), $I_0$ (instead of $\Lambda$), $\epsilon$. Three other
parameters are directly taken from phenomenology:
$g_A \simeq 0.8$
\ci{20},
$\hat m_q (1 {\rm GeV}) \simeq 7$ MeV and $B_0 \simeq 1.4$ GeV \ci{13,21}
($B_0$ instead of $c_1$). The two last ones provide the pion mass
$m_\pi \simeq \sqrt{2 m_q B_0} \simeq 140 $ MeV. One more parameter,
$\Delta_{SP}$ is related to the four-quark condensate which can be estimated
as in \gl{CSR} by employing the vacuum dominance hypothesis. However this
parameter makes almost no influence on the main meson characteristics
(as it follows
from \gl{cons2}, \gl{L-8} the deviation is of order 2\%). 
Therefore we neglect it in the forthcoming fit. 

The above 7 unknown parameters are involved in the determination of 7
physical quantities: 
$m_\sigma, m_\Pi, F_0, F_\Pi, F_\sigma, L_5, L_8$, but they
are connected by 3 relations; namely, the mass-gap equation \gl{dena}
and the two CSR constraints \gl{beta} and \gl{cons2}. Thus
the set of physical entries is  sufficient in principle to find all the
ECQM constants from  phenomenology and make some non-trivial predictions.
In a way this is a little miracle since we started with ${\cal O}(50)$
operators, each one with an unknown constant. In addition our model
reproduces the same value for the chiral constants
$L_1, L_2,L_3,L_9$ and $L_{10}$ as
the simplest chiral quark model\ci{2} in what concerns the
radiatively induced quark contribution.
These cannot be directly read from
\gl{log} because we have
retained there only the logarithmically enhanced
terms. It is known, however, that the pieces in the effective action
leading to the above $L_i$'s, albeit not logarithmically enhanced, are
universal and independent of the regulator. Of course the direct
bare contribution to the $L_i$'s is not included in our approach
(see appendix C).

Let us perform a
tentative fit using the phenomenological values for the heavy pion mass
$m_\Pi \simeq 1.3$ GeV \ci{10} and for
$F_0 \simeq$ 90 MeV \ci{13}. First
we relate the scalar meson mass to the chiral constant $L_8$ by means of
\gl{L-8}
\be
m^2_\sigma = m^2_\Pi(
\frac{16 L_8 m^2_\Pi }{ F_0^2} -1)^{-1}. \la{msig}
\ee
When taking the above input for  $m_\Pi, F_0$ and the phenomenological
value of $L_8 = (0.9 \pm 0.4)\times 10^{-3}$ \ci{13} one obtains
the mean value for the $\sigma$ meson
mass $m_\sigma \simeq 0.9 $ GeV which is well compatible with the
experimental data \ci{10}. The central value corresponds to
$|\beta|\simeq 0.7$.

Next let us use the definition of
the pion decay constant \gl{fpi} in order to eliminate the parameter $I_0$
\be
\sqrt{N_c I_0} = \frac{F_0 \sqrt{1 - \delta^2}}{g_A |x| M_0}. \la{I-0}
\ee
After substituting in \gl{denot} one obtains
\be
|\beta| = \frac{B_0 g_A \bar g_S x}{\sqrt{1 - \delta^2} M_0}. \la{beta1}
\ee
This relation can be combined with the definition of the chiral
constant $L_5$ \gl{L5} and with eq.\gl{scps} for the scalar meson mass
reexpressed in terms of $x, M_0, \bar g_S$
\be
L_5 = \frac{F_0^2(1 - \delta^2)}{8 B_0 M_0 (3 - 2x)},\qquad
m^2_\sigma =  2 M_0^2 \frac{3 - 2x}{\bar g_S x}. \la{L5ms}
\ee
Evidently there are two combinations of eqs.\gl{beta1},
\gl{L5ms} which do not
depend on $M_0$
\ba
&&\sqrt{1 - \delta^2} = \frac{4 L_5 |\beta| m^2_\sigma}{g_A F_0^2}
= \frac{4 L_5 m^2_\Pi |\beta|(1-\beta^2) }{g_A F_0^2 },
\la{delt}\\
&& \rho\equiv \bar g_S x (3 - 2x) = \frac{8 L_5^2 m^6_\Pi
\beta^4 (1-\beta^2)^3}{g_A^4 F_0^4 B_0^2 }
 . \la{kappa}
\ea
The product $\bar g_S x$  is positive for solutions delivering a true
minimum of the effective potential both in the
subcritical and  overcritical regimes
(see the discussion in
section 4). Hence $x < 3/2$.
For the central value of $L_5 = (2.2 \pm
0.2)\times 10^{-3}$ \ci{22} one gets
\be
\delta \simeq 0.56, \qquad \rho \simeq 0.12 \,. \la{num1}
\ee
We now eliminate $\bar g_S$ from the mass-gap eq.\gl{dena} in favour of
$\rho$
\be
(2 - \rho) x^2 - 5x + 3 = 0,
\ee
finding
\be
x_{\pm} = \frac{5 \pm \sqrt{1 + 12 \rho}}{4 - 2 \rho} .\la{solu}
\ee
Clearly if $\rho < 2$ then both solutions are positive but only the second
one, $x_{-} < 1$, delivers a true minimum. It corresponds to the subcritical
regime with the positive $\bar g_S$. The first solution $x_{+} > 3/2$
does not correspond to an absolute minimum. In the QCD case
we have estimated  that $\rho \simeq 0.12$ and
therefore the ECQM effective action motivated by QCD is realized in
the subcritical domain. This is one of our main conclusions:
\be
x \simeq 0.92, \qquad \bar g_S \simeq 0.11 > 0,\qquad \beta > 0 .
\ee
Of course the fact that $\bar g_S \ll 1$ does not by itself imply
that the coupling constant associated to the
NJL-type operators is therefore relatively weak as $\bar g_S$ is subject
to an arbitrary normalization. However, we have already argued when
discussing the gap equation that the normalization used here is the natural
one. Indeed that the theory is in a weak coupling regime can be seen in that
$\Sigma_0$ is close to  $M_0$.

Using eq. \gl{beta1} one finds the central values for the bare constituent
mass, $M_0 \simeq 190$ MeV (i.e. $\Sigma_0 \simeq 180$ MeV)
and consequently $I_0 \simeq 0.1$; that is
$\Lambda \sim 1.4$ GeV, in very good
agreement with the naive estimate from ChPT, $\Lambda\simeq 1.2$ GeV.
For the above values of $B_0, x, M_0, I_0$
one finds from \gl{cond1} that $c_1 \simeq - (185 {\rm MeV})^3$.

Using the relation for the heavy pion mass \gl{scps}
we can estimate the asymmetry between scalar and pseudoscalar four-fermion
interaction
\be
\gamma = \bar g_S \left( \frac{m^2_\Pi \delta^2}{2 M_0^2} - x ^2\right)
\simeq 0.7 .
\ee
Thereby the pseudoscalar four-quark coupling, $\bar g_P \simeq 0.16$
 has a tendency to be stronger than
the scalar one. However the pretty large errors in the determination of
the phenomenological coupling constants $L_5, L_8$ makes this conclusion
somewhat qualitative.

Finally we apply eq.\gl{beta} to
fix the parameter $\epsilon$. According to \gl{beta} it may have two
values: $\epsilon \simeq - 0.14$ or $\epsilon \simeq - 0.7$. The former one
is close to zero, compatible with a
 hidden left-right symmetry of the ECQM
action
\footnote{The exact symmetry is realized for a scalar meson
with a mass $\sim 650 $ MeV}, whereas the latter one is
quite compatible (within the error bar)  with the possibility to have only one
type of the mass term, $\epsilon = \pm 0.5 $ .

After substitution into the definition \gl{fPi} of
the weak decay constant of
the heavy pion $F_\Pi$ one arrives at the prediction
\ba
d_1=\frac{F_\Pi m^2_\Pi}{F_0 m^2_\pi} &=&
\frac{\sqrt{1 - \delta^2}}{\delta}
\left( \frac{2\epsilon\gamma M_0}{\bar g_S x g_A B_0} + 1 \right)\no
&=& \pm \frac{\sqrt{1 - \beta^2}}{\beta}
\simeq   \pm 1.0 \la{fPi1}
\ea
The constant $F_\Pi$ is not yet measured experimentally
but has been estimated
recently using modified sum rules \ci{22} to correspond to
$|d_1| = 2.2 \pm 0.8$. Finally, we estimate the scalar
meson decay constant $F_\sigma\simeq 1.4 F_\pi$, which
quantifies the deviations from the predictions of
the linear sigma model.

By now we have derived the following three predictions:
$m_\sigma$ (for a given value of $L_8$, or the other way round),
$F_\Pi$ and $F_\sigma$. We have so far considered the central
values. Let us now examine the error bars. A conservative estimate
$400 {\rm MeV}< m_\sigma < 1200 {\rm MeV}$ is given in
[11]. From \gl{L-8} it leads to $0.7\times 10^{-3} < L_8 <
3.5\times 10^{-3}$, to be confronted with the chiral dynamics
estimate $L_8=(0.9\pm 0.4)\times 10^{-3}$. We see that too
light values of $m_\sigma$ are ruled out from pion and kaon physics
since the allowed range for $L_8$ leads to
$700 {\rm MeV}< m_\sigma < 1200 {\rm MeV}$.
Conversely, one sees from \gl{L-8} that $L_8\ge 0.6\times 10^{-3}$
Then, assuming compatibility of these phenomenological
data, we derive
\be
L_8= (1.0\pm 0.3)\times 10^{-3}, \qquad
700 {\rm MeV} < m_\sigma < 1200 {\rm MeV}.
\ee
For these ranges we find from \gl{fPi1} and \gl{Fsig}
\be
|d_1|= 0.6 - 2.5, \qquad
F_\Pi = (0.4 - 1.8)\times 10^{-2} F_\pi, \qquad
F_\sigma = (1.2 - 2.6 ) F_\pi.
\ee
These quantities have not yet been measured. It is worth pointing out
that they are sensitive to the sigma meson mass and therefore
constitute a good laboratory for improving our knowledge of
the scalar channel.

We have seen  that all parameters of the ECQM can be derived from
meson phenomenology at low and intermediate energies and that this model
is predictive. It is quite remarkable that the values we obtain for
the different coefficients and constants of the effective action are
in excellent agreement with our theoretical expectations. For instance,
$f_0$ is smaller but of the same order as $F_0$; the coupling constant
$g_{P1}$ is slightly bigger, but again of the same order as $g_{S0}$,
as some theoretical considerations suggest (see appendix B). The chiral
symmetry breaking scale $\Lambda$ comes out naturally in the expected range,
while one of the allowed values of
the parameter $\epsilon$ is compatible with the
left-right symmetry discussed in the text.

\bigskip

\noindent{\bf 9. Conclusions}

\medskip

Let us summarize  the key points of our approach, outline the main
results and
explore possible continuations of our work

1) We have seen that the original chiral quark model is not sufficient to
describe  meson physics in the GeV energy region. The
structural constants $L_5, L_8$ of the effective chiral lagrangian also require
four-quark interaction operators  to be adequately accounted for.
Moreover it can be shown (see appendix B) that a consistent QCD
bosonization,
with the help of bosonic chiral fields incorporated into the constituent
mass operator,  generically  induce  four-quark interaction of
the NJL type.
Thus the most notable features of the ECQM derive quite naturally
from  QCD.

2) We have proposed a very general approach based on the
wilsonian ideas of retaining only the light degrees of freedom
and using symmetry and dimensional considerations. Thus, the
ECQM should not be regarded as `just another model'
 for low energy QCD,
but rather as a genuine and correct parametrization of
QCD below the CSB scale $\Lambda$ in the scalar and pseudoscalar
channels, much
in the same way as the chiral lagrangian is not a model of QCD
below the
meson threshold, but QCD itself.

3) Two novel ingredients, namely the matching to
QCD, via the chiral symmetry restoration at high energies,
and the
renormalization group analysis play a crucial role in eliminating some
of the unknown parameters and setting the accuracy of our approach.

4) We have made some approximations such as neglecting
quark operators of canonical dimension  $ d  > 6$ and purely chiral
operators with four derivatives of chiral dimension $d \geq 4$,
the large-$N_c$ approach and the large-log approximation.
These systematic approximations have obviously some bearing
on the accuracy of the present approach to the
ECQM, but it is hard to go beyond them
without more precise knowledge
of the value of the
bare couplings at the scale $\Lambda$, something which is
at present impossible
without solving QCD (perhaps by performing numerical
simulations in the ECQM and
matching them with QCD\ci{rebbi}). At any rate we do not really
expect the  uncertainty of the present approach
to be worse, at comparable energies, than
the one characteristic of ChPT.

5) The approximations we have made in our action
are theoretically self-consistent to ensure renormalizability of
the low-energy action. It is
remarkable that
 the number of parameters which have been
retained happens to be sufficient
to fit well the phenomenology of $\pi, \Pi$
and $\sigma$ mesons and this with
quite natural values as dictated by dimensional and symmetry
considerations.
The model allows to  reliably estimate the scalar meson masses
and decay constants from other data. They
come out naturally in a range which is not in obvious conflict
with low energy phenomenology, something that does not happen in the
NJL model.

6) Our analysis shows that the incorporation of the
chiral constituent mass $M_0$, as in the conventional CQM, leads to a
relatively small value for the coupling constant
of the effective NJL-like
four fermion interactions, which come out to be relatively weak.

7) It is remarkable that the scalar and pseudoscalar fermion
self-interaction
do not appear to be equal. This is because chiral symmetry
is realized in low energy QCD in a rather subtle and not totally
manifest way at the level of the effective action.
It is worth pointing out also that the structure of soft
breaking mass terms is
compatible with the one corresponding to the $L\leftrightarrow R$
symmetry in eq. (136).

8) The  generalization to three light
quarks is in progress.
One adds then two more 4-fermion operators
to the list \gl{4f1} - \gl{4f4}, but gains the physics of the scalar
and pseudoscalar $SU(3)$-octet mesons.

9) In this spirit we propose the pseudoscalar singlet state to be
the natural heavy partner of the $\sigma$ meson. Thereby
the resolution of
the $U_A(1)$ problem and the $\eta'$ meson
mass will be implemented in the ECQM
 by means of 4-quark flavour-singlet pseudoscalar forces. Obviously one
should go away from the large $N_c$ limit to study this extension.

10) The generalization to include scalar isotriplet or spin $\geq 1$
 quark currents 
is quite straightforward. The effects of such
resonances can be
partially absorbed in the structural constants of the ECQM, for instance,
in $\delta g_A$. Hence  their incorporation will modify also
the latter ones
(a related discussion in the case of extended NJL model can be found in
\ci{6,20}).

\bigskip
\noindent
{\bf Acknowledgements}
\medskip

This work is supported by grants CICYT  AEN95-0590 and CIRIT
1996SGR00066.
A.A. is
supported by the Spanish Ministerio de Educaci\'on y Cultura
(Ref. SAB1995-0439)
and partially by   grants RFFI 98-02-18137,\
INTAS-93-283-ext
and GRACENAS 6-19-97. We are grateful to J.Soto for useful
discussions and comments. D.E. wishes also to thank E. Aguirre
for inspiring comments.

%\vspace{0.5cm}
\newpage

\noindent{\bf Appendix A}

\medskip

Let us justify here the use of the EOM  \gl{em1} and \gl{em2}
to eliminate certain higher-dimensional
vertices in section 3. For simplicity we do not introduce
external sources. The simpler case concerns the EOM of
quark fields \gl{em2} as they are linear. The CQM leads to the
generalized Dirac operator
$$
i \hat D_{CQM} = i\not\!\partial  -  M_0 (U^+ P_R + U P_L), \eqno{(A1)}
$$
acting on quark fields $q(x)$. Since we eventually integrate out the
quark fields, we need to use the field equation
 $\hat D_{CQM}q(x) = 0$ inside the fermion loop to eliminate the unwanted
operators. We shall treat all the operators with dimension $\ge 4$ as
perturbations and thus consider the v.e.v. of products of them, performing
the appropriate Wick contractions. The CQM Dirac derivative
will thus act on the fermion propagator
$$
\underline{q(x) \bar q(y)}\equiv\langle T(q(x) \bar q(y))\rangle
 =
\langle x | \frac{1}{\hat D_{CQM}}| y\rangle .  \eqno{(A2)}
$$
giving
$$
\hat D_{CQM}\,\underline{q(x) \bar q(y)} = \delta^{(4)} (x - y) \mbox{\bf I}.
\eqno{(A3)}
$$

The Dirac derivative $\not\!\partial$ appears in operators of $d \geq 5$
like \gl{d51},\gl{d58},\gl{d59},\gl{d61},\gl{d62} etc. The general structure
 is
$$
\int d^4x \bar q(x) O_n\! \not\!\partial q(x) \qquad
\mbox{or }\qquad \int d^4x \bar q(x)\! \not\!\partial O_n q(x), \eqno{(A4)}
$$
where $O_n$ is a differential operator of finite order
and dimension $n\ge 1$
which contains also
the relevant matrix structure, in particular, left or right projectors if
necessary. Thus one gets, for instance,
$$
\int d^4x \int d^4y \,\bar q(x) O_n \!\not\!\partial
\underline{q(x)\bar q(y)}  O_m q(y) = \eqno{(A5)}$$
$$- \int d^4x \int d^4y\, \bar q(x) O_n  M_0 (U^+ P_R + U P_L)
\underline{q(x)\bar q(y)} O_m q(y)
+ \int d^4x \, \bar q(x) O_n
 O_m q(x) $$
The last operator in (A5) has dimensionality
$d=
n + m + 3 \geq 5 $. Thus the application of the EOM generates
operators of at least the same dimension without the
Dirac derivative (or, in general, with one Dirac derivative less).
The result of the application of the EOM is
the removal
of all Dirac derivatives acting on quark fields at the
expense of modifying the coupling constants in operators of $d\geq 5$
without  derivatives. In particular, the $d = 5$ operators are mapped
into a set of $d \geq 5$ operators,  the $d = 6$ vertices are mapped
into a set of $d \ge 6$ operators, etc.

One can improve this reduction when dealing with the ECQM
effective action
\gl{ECQM}. Indeed, the latter one includes all vertices of $d
\leq 4$ and therefore the higher dimensional operators
start now from
$d = 5$. It corresponds to $m \geq 2$ in (A5).
Thus in the ECQM, with the help of
the EOM, the $d = 5$ vertices are mapped
onto a set of $d \geq 6$ vertices etc.

In an analogous way one can show that the use of the EOM for chiral
fields \gl{em1} leads also to the modification of coupling constants
in higher dimensional vertices without an insertion
of the operator $\partial_\mu l^\mu$. The proof is now complicated by
the nonlinear structure of the EOM. Nevertheless one can expand the
chiral variables $U$ in powers of $\pi$ and
analyze the pertinent EOM reduction in ChPT.
The
ChPT counting rules assign dimension zero to the
pion field, and $\partial_\mu l^\mu \sim
\partial^2 \pi$ thus carries chiral dimension $d = 2$.
The equations of motion can  clearly be applied to external
pion fields without further ado. As for those operators contributing to
internal
pion loops they will give rise, after the appropriate contractions, to the
pion propagator
$$
\left\langle T(\pi(x)\pi(y))\right\rangle =
\langle x | \frac{1}{\partial^2 }|y\rangle. \eqno{(A6)}
$$
The
application of the Klein-Gordon operator $\partial^2$ arising
from the EOM
to the above Green function  gives
$$
\partial^2 \left\langle T(\pi(x)\pi(y))\right\rangle
= \delta^{(4)} (x - y) \mbox{\bf I}.  \eqno{(A7)}
$$
Since all operators of $d \leq 3$ are included in the effective
lagrangian of the CQM the set of higher dimensional operators treated
as perturbations consists of those ones with $d \geq 4$. Hence
after the EOM reduction in two vertices connected by one
pion propagator one expects that additional local
vertices of $d \geq 4 + 4 - 2 = 6$ are induced similarly to (A3).
Evidently these $d=6$ operators have also chiral invariant structures and
therefore represent particular combinations of the complete set of
$d=6$ operators classified in section 3. If some of them still contain
insertions of the operators $\partial_\mu l^\mu$ and/or $\not\!\partial$
acting directly on quark fields, repeated use
of the EOM produces additional
operators of $d \geq 6$ without any such insertions.

\bigskip

\noindent{\bf Appendix B}
\medskip

Below the chiral symmetry breaking scale $\Lambda$, the (in
the chiral limit) massless Goldstone bosons,
associated to the local chiral phase of
the quarks,  appear. To avoid overcounting we must extract the chiral phase
from the QCD partition function. This naturally modifies the QCD
action. Let us see how and that, in particular,
the separation of chiral fields in QCD is in general
accompanied with the pertinent four-quark interaction. We
separate the local chiral phase by
 exploiting the following functional identity
(for a related approach see \ci{slavnov})
$$
1 =  \Delta \int {\cal D}U \,
 \prod_x  \delta \left(
\bar q_R U q_L(x) + \bar q_L U^\dagger q_R(x)\right)   , \eqno{(B1)}
$$
where $U$ are $SU(2)$ fields and  $\Delta$ is the
required Faddeev-Popov determinant.

Now we proceed to calculate the determinant and use the conventional
mapping of chiral $SU(2)$ fields onto $S^3$
$$
U = n^0 \mbox{\bf I} + i \tau^a n^a,\quad a = 1,2,3\qquad
(n_k)^2\equiv (n^0)^2 + n^a n^a  = 1, \eqno{(B2)}
$$
with the notation $k = 0,1,2,3$.
Then the properly normalized integral over chiral fields is
evaluated as follows
$$
\int {\cal D}U   \, \prod_x  \delta \left(
\bar q_R U q_L(x) + \bar q_L U^\dagger q_R(x)\right)
= \int \prod_x  \frac{d^4 n(x)}{C} \delta((n_k(x))^2 - 1)\,
\delta(\mbox{\bf J}_k(x) n_k(x))$$
$$ = \prod_x \,
\left(\mbox{\bf J}_k(x)\mbox{\bf J}_k(x)\right)^{- 1/2},
\eqno{(B3)}$$
where $C$ is a normalization constant and the quark currents are
$$
\mbox{\bf J}_0(x) \equiv \bar q q(x);\qquad \mbox{\bf J}_a(x) \equiv
i\bar q \gamma_5 \tau^a q(x). \eqno{(B4)}
$$
Therefore the determinant supporting the identity (B1) reads
$$
\Delta = \prod_x \left((\bar q q(x))^2 -
(\bar q \gamma_5 \tau^a q(x))^2\right)^{1/2}. \eqno{(B5)}
$$
This four-quark operator can be exponentiated
with the help of auxiliary fields. The $\delta$ function
in (B1) can be exponentiated as well by using the functional
Fourier representation. All together allows us to write
$$
1 =  \int {\cal D}U\,{\cal D}\mu \,{\cal D}\phi\,{\cal D}\bar c
\,{\cal D}c  \,
 \exp\int d^4x\,(i\mu(x)(
\bar q_R U q_L(x) + \bar q_L U^\dagger q_R(x)) $$
$$ -
(\phi^2(x) + \bar c c(x)) ((\bar q q(x))^2 -
(\bar q \gamma_5 \tau^a q(x))^2 )), \eqno{(B6)}
$$
where $U(x), \mu(x), \phi(x)$ are
bosonic fields and $\bar c(x), c(x)$  are
scalar fermionic ones (ghost fields).

After insertion of this identity in the QCD generating functional
one can argue that the quantum average over the gluon vacuum should
generically induce (since they are not protected by any symmetry)
non-zero v.e.v. for the auxiliary fields
$\langle \mu \rangle \equiv M_0$ and
$\langle\phi^2(x) + \bar c c(x)\rangle
\equiv g/4N_{c} \Lambda^2 $. Should all the above v.e.v. be equal to
zero we would conclude that the chiral symmetry is not broken and the $U$
field would entirely decouple. Although we cannot at present demonstrate
the actual mechanism, we know that this is not the case in low energy QCD
and from that point on the analysis presented in this paper takes over.

\bigskip

\noindent
{\bf Appendix C}

\medskip

Here we shall compare our results
with those obtained in the pure CQM after
inclusion of gluonic corrections and see how they fit together.
In \ci{2} the values for $F_0$ and the low energy chiral
constants $L_i$ were derived considering both the contribution
which is radiatively induced by the fermion loop with constituent
mass $M_0$
and additional gluon
contributions expressed in terms of v.e.v. of gauge
invariant gluon condensates. The leading gauge invariant operator
is naturally $\langle \alpha_s GG\rangle$ and this was the only
one explicitly considered in \ci{2}.

As discussed in the text, the chiral constants $L_1$, $L_2$,
$L_3$, $L_9$ and $L_{10}$, which in the CQM get a contribution
from the constituent quark loop
$$
8L_1=4L_2=-2L_3=L_9=-2L_{10}=\frac{N_c}{48\pi^2},
\eqno(C1)$$
are exactly reproduced by the ECQM. The gluonic corrections
obtained in \ci{2} would correspond to what in this paper we call
bare contributions (and which we have
explicitly considered here only for $F_0$). For the present
discussion we take $g_A=1$, just as the authors of \ci{2} did.
We have
obtained the following expression for $F_0$
$$
F_0^2= N_cM_0^2 I_0 +f_0^2. \eqno(C2)
$$
With the same notations
the following result was obtained in \ci{2}
$$
F_0^2= N_cM_0^2 I_0 + \frac{1}{24\pi M_0^2} \langle \alpha_s GG\rangle,
\eqno(C3)$$
and
 the identification of $f_0$ is obvious. Notice that
the scale $\mu$ which was undetermined in \ci{2} is 
to be identified with the CSB one $\Lambda$. We know this because
we have good control on the renormalization group behaviour in
the low energy effective theory. We see clearly that
the bare coefficients contain all the gluonic contributions which
accumulate all gluonic modes both above and below $\Lambda$
which cannot be expressed in terms of the constituent mass $M_0$.

As to the chiral constants $L_5$ and $L_8$, using
the large log approximation, we read from
\ci{2}
$$
L_5= \frac{N_c M_0}{8 B_0} I_0, \eqno(C4)
$$
and
$$
L_8=\frac{N_c M_0}{16 B_0} I_0, \eqno(C5)
$$
both of which can be made to agree with our results in the 
$g_{S0} = g_{P1} =0$ limit for $\delta f_0 = 0$.
This limit is quite straightforward to take in $L_5$ (see
\gl{L5}), but it is worth recalling that in the present approach
$L_5$ is generated through $\sigma$ exchange and then the corresponding
amplitude has the correct high energy limit. The situation for
$L_8$ is a bit more subtle. The limit in \gl{L-8} cannot be taken
naively in the final expression for $L_8$ which vanishes, 
but rather we must replace $Z_\sigma$ and $Z_\Pi$
by their expressions in the first term in \gl{L-8} and then expand in
powers of $g_{S0} = g_{P1}$. We then recover $(C5)$.
The reason for this subtlety
must be eventually traced back
to the CSR constraints.
On the other hand, in \ci{2}, and also
using the leading log approximation,
one gets
$$
C_q= N_c M_0^3 I_0  \eqno(C6)
$$
This shows that within this approximation $M_0$, and consequently
$L_5$,  must be negative, in contradiction with experiment.
In \ci{2} this difficulty is circumvented due to the presence of
(unfortunately non-universal) finite parts which contribute to
$L_5$ with the opposite sign and eventually
allow getting around this
sign problem, although
the final fit for $L_5$, $L_8$ is not brilliant. We
have seen in the main
text how in the ECQM the renormalization group implies, on the one hand,
the presence of some
logarithmically enhanced contact terms which make
possible for $C_q$ and $L_5$ to have opposite signs without having
to appeal to finite non-universal constants.

\end{document}